\documentclass{aa} 
\usepackage{txfonts}
\usepackage{graphicx}
\usepackage{natbib}

\bibpunct{(}{)}{;}{a}{}{,}
\begin{document}

\title{Two-component magnetohydrodynamical outflows\\ around young stellar objects} 
\subtitle{Interplay between
  stellar magnetospheric winds and disc-driven jets}

  \titlerunning{Two-component MHD outflows around YSO}
  \authorrunning{Meliani, Casse \& Sauty}
  \author{Z. Meliani
           \inst{1,2,3}
           \and  F. Casse
           \inst{2}
           \and  C. Sauty
           \inst{1}
          }

   \offprints{Z. Meliani}

\institute{  
Observatoire de Paris, L.U.Th., F-92190 Meudon,
           France\\
 \email{zakaria.meliani@obspm.fr, Christophe.Sauty@obspm.fr}
    \and AstroParticule \& Cosmologie (APC)\thanks{UMR 7164 (CNRS,
      Univ. Paris 7, CEA, Observatoire de Paris)} - Universit\'e Paris 7, 11
    place Marcelin Berthelot, F-75231 Paris Cedex 05, France\\
 \email{fcasse@apc.univ-paris7.fr}
 \and Max Planck Institute for Astrophysics, Box 1317, D-85741 Garching, 
  Germany}

   \date{Received ... / accepted ...}

   \abstract{We present the first-ever simulations of non-ideal
     magnetohydrodynamical (MHD)
     stellar magnetospheric winds coupled with disc-driven jets where the resistive and
     viscous accretion disc is self-consistently described.}{ These
     innovative MHD simulations are devoted to the study of the interplay
     between a stellar wind (having different ejection mass rates) and an
     MHD disc-driven jet embedding the stellar wind.}{The
     transmagnetosonic, collimated MHD outflows  are investigated 
     numerically using the VAC code. We first investigate the various
     angular momentum transports occurring in the magneto-viscous accretion
     disc.  
     We then
     analyze the modifications induced by the interaction between the two
     components of the outflow.}{Our simulations show that the inner
     outflow is accelerated from the 
     central object's hot corona thanks to both the thermal pressure  and the
     Lorentz force. In our framework, the thermal acceleration is sustained 
     by the
     heating produced by  the  dissipated magnetic energy due to
     the turbulence. Conversely, the  outflow
     launched from the resistive accretion disc is mainly accelerated by
     the magneto-centrifugal force.}{The simulations show that the MHD 
     disc-driven
     outflow extracts angular momentum more efficiently  than do viscous
     effects in near-equipartition, thin-magnetized discs where turbulence is
     fully developed.  We also show that, when a dense inner stellar wind
     occurs, the resulting disc-driven jet has a different structure,
     namely a magnetic structure where poloidal magnetic field lines are
     more inclined because of the pressure caused by the stellar wind. This
     modification leads to both an enhanced mass-ejection rate in the
     disc-driven jet and a larger radial extension that is in
     better agreement with the observations, besides being more consistent.} 


\maketitle

\section{Introduction}

  Accreting stellar objects are often associated with collimated jets 
or  winds from accretion  discs. Most of those objects also show evidence
of winds 
originating in  a corona surrounding the central object. 
These accretion-ejection phenomena are observed in different astrophysical
sources ranging from young stellar  objects (YSOs), X-ray binaries,
planetary nebulae to active galactic nuclei (AGNs) (see, e.g. \citep{Livio07} 
and references therein). The outflow provides an efficient way of 
extracting  angular momentum and converting  gravitational energy 
from the accretion disc or from the central object into outflow kinetic energy.

Observations show that most of the jets are launched very close to the central engine.
In the case of YSOs, there is direct observational evidence
\citep{Burrowsetal96}
 as well as in the case of some microquasars 
\citep{Fenderetal97, Mirabel03}. For instance, it has been suggested that, in 
 microquasars, the fastest components of the outflow are
launched in the vicinity of the black hole \citep{meier03}.
Another piece of evidence that the outflow may originate in a region relatively 
close to the central  object is that the observed asymptotic velocity of the jet
is close to the escaping speed from the central engine. Thus there is a 
direct relation between the  asymptotic speed and the depth of the gravitational
potential \citep{Mirabel99, Livio99, Pringle93}. 
Moreover, the high-resolution images of $H_{\alpha}$ and $[OI]$ 
\citep{Bacciottietal02} reveal a continuous transverse variation of the 
jet's velocity, where the fastest and densest components are closer to the
central  axis.\\
The high velocity of the observed jet 
in YSOs suggests that they originate in a region that is no larger than
one astronomical unity (AU) in 
extent \citep{KwanTademaru88} and between 0.3 to  4.0 AU from the star in the
case of the LVC of DG Tau \citep{Andersonetal03}. 
  This theoretical prediction is supported, in the case of a disc wind,
by  observations of the rotation of several jets associated with TTauris 
\citep{Coffeyetal04}.  
Moreover, in the case of classical TTauris (CTTS),  
UV observations \citep{Beristainetal01,Dupreeetal05} reveal the presence of a 
warm wind whose temperature  is at least of $3\times 10^5 {\rm K}$.
It appears that the source of this wind is restricted to the star itself, where 
X-ray observations support the presence  of a hot 
corona in CTT stars \citep{Feigelson&Montmerle99}.
  These observations also suggest the existence in CTTS of stellar winds 
comparable to the solar wind. These winds may be both thermally and
magneto-centrifugally accelerated.

Since the discovery of the existence of winds and jets in astrophysics, enormous
progress has been made regarding the understanding of these phenomena.
At the same time, we still do not know precisely how the wind from the
central corona of the star or the compact object interacts with
the disc outflow and the  respective roles and differences between these two 
types of flows.\\
Several works have analytically and numerically studied the
formation of outflows launched from the accretion disc  \citep{BP82,
Cao&Spruit94, ContopoulosLovelace94, Ustyugovaetal95,
Ouyedetal97, VlahakisTsinganos98, CasseFerreira00, Casse&Keppens2002,
Casse&Keppens2004, Andersonetal05, Pudritzetal06}. Other
works have focused on the
outflows  from the hot  corona of the central objects  \citep{Sakurai85,
SautyTsinganos94, Fendt03,  Matt&Balick04}. \\
 In models dealing with
outflows launched from
accretion discs, the magnetic field plays a  key role  in the
accretion, the acceleration, and the collimation of the associated vertical
wind, which is also supported by recent observations  \citep{Donatietal05}. 

The detection of rotation signatures in TTauri jets gives extra
strong support to the magneto-centrifugal launching from the accretion 
disc. However in stellar ouflows, the wind must be thermally accelerated 
because of the strong heating of viscous and non-ideal magnetohydrodynamical
(MHD) mechanisms.  
 This acceleration increases and approaches the
magneto-centrifugal acceleration at least near the polar  axis. 
Some models have already investigated diffusive disc-driven jet launching. 
In some simulations the accretion disc was considered as a fixed, 
time-independent boundary condition, while a constant magnetic resistivity
prevails through the entire outflow \citep{Fendt&Cemeljic02}. \citet{Kuwa05}
have included an accretion disc in their resistive
simulations but only considered an uniform magnetic resistivity everywhere in
their computational domain. Because of the very different physical conditions
prevailing in the disk, the jet, and the external medium, it seems very unlikely
that this kind of resistivity is physically relevant. 
Models involving two component bipolar outflows have been proposed in the case 
of AGN as such as in Sol et al. (1989) and  Renaud \& Henri (1998), where an 
electron-positron central wind 
component is surrounded by an external, ideal MHD disc-driven jet. Another
two-component outflow model has been proposed by  
\citep{Bogovalov&Tsinganos05} where a relativistic pulsar wind is
surrounded and
collimated by an ideal MHD disc-driven wind.
 In the case of YSOs, two-component models were considered, such as  in
 X-wind outflows  \citep{SautyTsinganos94, 
Ferreiraetal00}. The inner component extracts its energies from the corona
around the central region (the central object and the inner part of the accretion
disc where an  advection-dominated accretion flow may exist), while the second
component is  launched from the thin  accretion disc.\\
The aim of this paper is to investigate the formation of a two-component 
outflow around YSOs, one coming from the thin accretion disc and the other
one injected 
from the hot corona of the central star. This work is developed on the 
base of the  disc wind simulations of
Casse \& Keppens (2002, 2004) (CK02, CK04). 
Its goal is to study the influence of the stellar wind on the structure 
and the dynamics of the jet around YSOs. Furthermore, we investigate the
consequences of the  energy dissipation in the outflow close to the polar axis.
Before that, we present simulations of the outflow launched from 
a resistive and viscous accretion disc where magnetic Prandtl number  
(ratio of the anomalous viscosity to the anomalous resistivity) equals  
unity.  This is the first time that viscosity and resistivity are considered 
together in the disc and included in an MHD simulation involving both the  
accretion disc and the related jet. In 
previous works, the viscous accretion disc was examined without taking the disc 
wind into account or with an imposed internal structure 
\citep{VonRekowskietal06}, which does not enable the authors to study the 
complete angular momentum transfer. 
Thus, the first part investigates the relative role of angular-momentum 
transport by viscosity and by the outflowing plasma and its influence on 
the formation of the outflow from the disc. Then in a second part, we 
present the results of our simulations of ideal MHD outflows launched from
resistive, viscous, accretion discs surrounding the turbulent 
wind accelerated from the hot corona of the central star. 


\section{Ideal MHD outflows arising from resistive, viscous, thin accretion discs}

  In this section we present the simulations of axisymmetric MHD outflows
generated from thin accretion discs where for the first time viscosity is
taken into account, together with resistivity. We recall that turbulence 
is believed to generate both anomalous resistivity {\it and} viscosity, 
such that the turbulent magnetic Prandtl number, which is the ratio of 
the viscous to the resistive transport coefficients,  should be of order 
unity within flows where turbulence is fully developed
\citep{Pouquetetal76,Kitchatinov&Pipin94}. This is supposed to be the case of 
accretion discs and stellar winds. The presence of two braking torques inside
the disc may achieve different disc-jet configurations, since the angular 
momentum transport is modified by the presence of viscosity.

\subsection{MHD equations}

To get the evolution of such a disc, we solve the system of
time-dependent resistive and viscous MHD equations, namely, the
conservation of mass, 
\begin{eqnarray}\label{Eq_Continuity}
\frac{\partial \rho}{\partial t} =
-\vec{\nabla}\cdot\left(\rho \vec{v}\right),
\end{eqnarray} 
and momentum,
\begin{eqnarray}\label{Eq_Momentum}
\frac{\partial \rho \vec{v}}{\partial t} + 
\vec{\nabla}\left(\vec{v} \rho \vec{v} - \vec{B} \vec{B}\right) + 
\nabla\left(\frac{B^2}{2} + P\right) + \rho \vec{\nabla} \Phi_{\rm G} = 
\nonumber \\
- \vec{\nabla}\cdot\left(\eta_{v} \hat{\Pi}\right).
\end{eqnarray}
 We also consider the energy conservation governing the temporal
  evolution of the total energy density $e$,
\begin{eqnarray}\label{Eq_dEnergydt}
e = \frac{\vec{B}^2}{2} + \frac{\rho \vec{v}^2}{2} + \frac{P}{\gamma - 1} &+&
\rho \Phi_{\rm G} \ ; \\
\frac{\partial e}{\partial t} + \vec{\nabla} \cdot \left[ \vec{v} 
\left( e + P + \frac{B^2}{2}\right) - \vec{B} \vec{B} \cdot \vec{v}\right] 
&=&  \nonumber \\
\eta_m \vec{J}^2 - \vec{B} \cdot \left(\vec{\nabla} \times \eta_m \vec{J}\right)
&-& \nabla \left(\vec{v} \cdot\eta_{v} \hat{\Pi}\right) \nonumber
\end{eqnarray} 
where $\rho$ is the plasma density, $\vec{v}$ the velocity, $P$ the thermal
pressure, $\vec{B}$ magnetic field, $\gamma$  the specific heat ratio
($C_P/C_V=5/3$), and $\vec{J} = \vec{\nabla}\times B$ is the 
current density. In this set of MHD equations, the thermal pressure is
derived from all conserved physical quantities. To close the set
of equation, we have included a perfect
gas equation of state linking the thermal pressure and plasma density to the plasma
temperature $T$, namely $T=P/\rho$.
The gravitational potential is given by
\begin{equation}
\Phi_{\rm G}= - \frac{GM_{\star}}{\left(R^2 + Z^2\right)^{1/2}}.
\end{equation}
 Note that both resistivity
($\eta_m$) and viscosity ($\eta_v$) are taken into account in the MHD set of
equations. The viscous stress tensor is given by $\eta_{v} \hat{\Pi}=
-\eta_{v} \left(\left(\vec{\nabla} \vec{v} +\vec{\nabla}  
\vec{v}^{\rm T}\right) + \frac{2}{3} \hat{I}\left(\vec{\nabla}\cdot
\vec{v}\right)\right) $. The last equation provides the temporal evolution of the
magnetic field, namely the induction equation 

\begin{eqnarray}\label{Eq_Induction}
 \frac{\partial B}{\partial t} + \vec{\nabla} \left(\vec{v} \vec{B} - 
\vec{B} \vec{v}\right)= - \vec{\nabla}\times\left(\eta_m \vec{J}\right)\ .
\end{eqnarray} 
The local heating and torque in the accretion disc are generated by 
the magnetic resistivity and hydrodynamical 
viscosity occurring in the disc. We adopt  a magnetic Prandtl number 
$Pr = \frac{\eta_{v}}{\eta_{m}}=1$ in our simulations as it is reasonable on
physical grounds (see references above). 

\subsection{Initial and boundary conditions}

 The initial density profile and velocities are set as follows, 
\begin{eqnarray}
\rho(R, Z) &=& \rho_o\max\left[5\times 10^{-6}, 
\frac{R^{\frac{3}{2}}}{\left(R^2 + R_{0}^2\right)^{\frac{3}{4}}}\right]\nonumber\\
&\times& \left(5 \times 10^{-6}, \left\{1 - \frac{\left(\gamma -1\right) Z^2}{H^2}\right\}^{\frac{1}
{\left(\gamma-1\right)}}\right),\\
V_{R}(R,Z) &=& - V_om_{s} \frac{R_{0}^{\frac{1}{2}}}{\left(R^2 + R_{0}^2\right)^{\frac{1}{4}}}
\exp\left(-\frac{2 Z^2}{H^2}\right) = V_{Z} \frac{R}{Z}\\
V_{\theta}(R,Z) &=& V_0\left(1 - \epsilon^2\right) 
\frac{R_{0}^{\frac{1}{2}}}{\epsilon \left(R^2 + R_{0}^2\right)^{\frac{1}{4}}}
\exp\left(-\frac{2 Z^2}{H^2}\right)
\end{eqnarray}
where $H = \epsilon R$ is the disc height, which is proportional to the radius $R$, 
via the disc aspect ratio $\epsilon \sim C_s/V_K$ linking the disc sound speed
$C_s$ to the Keplerian velocity $V_K$. 
We deliberately choose $\epsilon$
smaller than unity to get a thin disc where thermal pressure
gradient is smaller than both centrifugal and gravitational forces
\citep{Wardle&Konigl93}. The parameter $m_{s} =0.1$  ensures that the
initial poloidal flow remains subsonic.\\
\begin{figure}[t]
\begin{center}
{\rotatebox{0}{\resizebox{4.5cm}{9cm}{\includegraphics{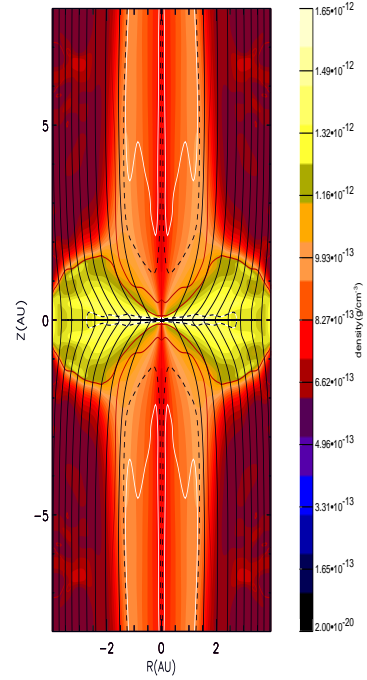}}}}
\caption{Density contours in the poloidal plane of an
  accretion-ejection structure where a viscous and resistive MHD disc is
  launching a collimated jet. Magnetic field lines are drawn 
in black solid lines, while the fast magnetosonic surface corresponds to the
white solid line (Alfv\`en surface is the black dotted line). 
The size of the sink region
   is $R_i=0.1 {\rm AU}$ and the stellar mass is $1 M_{\odot}$.}
\label{Fig1}
\end{center}
\end{figure}
The initial magnetic field configuration is taken as in CK04,
\begin{eqnarray}\label{Magn_prescription}
F(R,Z) &=& \sqrt{\beta_{p}} B_o\frac{R_{0}^{5/4} R^2}{\left(R_{0}^2 + R^{2}\right)^{\frac{5}{8}}} 
\frac{1}{1 + \zeta Z^2/H^2}\,,\nonumber\\
B_{R}(R, Z) &=& - \frac{1}{R} \frac{\partial F}{\partial Z}\,,\nonumber\\
B_{Z}(R, Z) &=&   \frac{1}{R} \frac{\partial F}{\partial R} + 
\frac{\sqrt(\beta_{Z})}{\left(1 + R^2\right)}\,,\nonumber\\
B_{\theta}(R,Z) &=& 0\,.
\end{eqnarray}
where $\beta_{p} = B^2/P$ is set to $0.6$  to ensure that the magnetic pressure
remain  on the order of the thermal pressure, a necessary condition for a
disc launching large-scale stable jets \citep{Ferreira&Pelletier95}. 
The parameter $\zeta$ controls the initial bending of
the magnetic surface. In the following simulation we set $\zeta = 0.04$.
The reader may shift from
dimensionless quantities used in our simulations to physical quantities by
setting the mass of the star $M_*$, the accretion rate $\dot{M}_{a}$, as
well as the size of the disc inner radius $R_i$,  
\begin{eqnarray}
\rho_{0} &=& 2.4 \times 10^{-12} \left(\frac{\dot{M}_{a}}{10^{-7}
  M_{\odot}yr^{-1}}\right)\left(\frac{M_{\star}}{M_{\odot}}\right)^{-1/2}\left(\frac{R_i}{0.1
  AU}\right)^{-3/2} {\rm g\;cm^{-3}}\nonumber \\
V_{o} &=& 9.5 \,\left(\frac{M_\star}{M_{\odot}}\right)^{1/2}\left(\frac{R_i}{0.1 AU}\right)^{-1/2} {\rm km/s}\nonumber\\
T_{0}&=& 10^{4}\left(\frac{M_\star}{M_{\odot}}\right)\left(\frac{R_i}{0.1
  AU}\right)^{-1}{\rm K}\nonumber \\
B_o&=&7\ \beta_p^{1/2}\left(\frac{\dot{M}_{a}}{10^{-7} M_{\odot}yr^{-1}}\right)^{1/2}\left(\frac{M_{\star}}{M_{\odot}}\right)^{-1/4}\left(\frac{R_i}{0.1
  AU}\right)^{-5/4} G
\end{eqnarray}

\noindent The expression of the anomalous resistivity in the accretion disc is
as in  CK04, 
\begin{eqnarray}\label{Eq_Resistivity_Disc}
 \eta_{m} &=& \frac{\eta_{v}}{Pr} = \alpha_{m} 
\left.V_{A}\right\vert_{Z = 0} H \exp\left(-2 \frac{Z^2}{H^2}\right)
\end{eqnarray}
The anomalous resistivity $\eta_{m}$ has the same origin as the turbulent 
viscosity since time correlations for hydrodynamic and magnetic turbulence
are the same. Thus we likewise introduce an anomalous viscosity $\eta_{v}$ 
equal to $\eta_{m}$. As in CK04, we have replaced in the $\alpha$-prescription 
the sound speed by the Alfv\'en speed because the
anomalous resistivity is related to the small-scale turbulent magnetic
field. Since the disc remains near equipartition between thermal pressure
and magnetic pressure, this does not affect the value of the transport
coefficients.
Through the dependence on the Alfv\'en velocity,
this becomes a  profile varying in time and space that essentially vanishes
outside the disc. We take $\alpha_{m}=0.1$ smaller than one to ensure that 
the Ohmic dissipation rate at the mid-plane of the accretion disc does not
exceed the rate of gravitational energy release \citep{Konigl95}. 
{ However, for a lower value of $\alpha_{m}<0.1$,  
the  energy released by accretion is insufficient to produce a strong
collimated wind crossing all critical surfaces. In the case of weak
resistivity, the resulting outflow remains weak, and the opening angle of the
jet is small \citep{Ferreira97}. This result has been confirmed by
numerical calculations \citep{Casse&Keppens2002}.}\\ 
\begin{figure}[t]
{\rotatebox{0}{\resizebox{8cm}{6cm}{\includegraphics{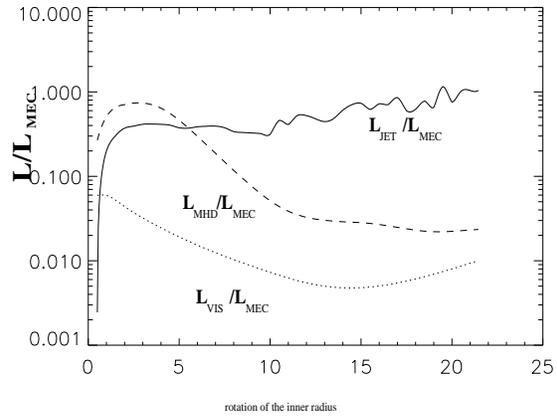}}}}
\caption{Temporal evolution of several angular momentum fluxes occurring
  inside the accretion-ejection structure displayed in
  Fig.\ref{Fig1}. The various fluxes are normalized to
  the amount of angular momentum removed from the disc $L_{MEC}$. The
  dominant way to remove disc angular momentum is provided by the magnetic
  torque leading to the creation of a jet.}  
\label{Fig2}
\end{figure}
\begin{figure}[ht]
\begin{center}
{\rotatebox{0}{\resizebox{4.5cm}{9cm}{\includegraphics{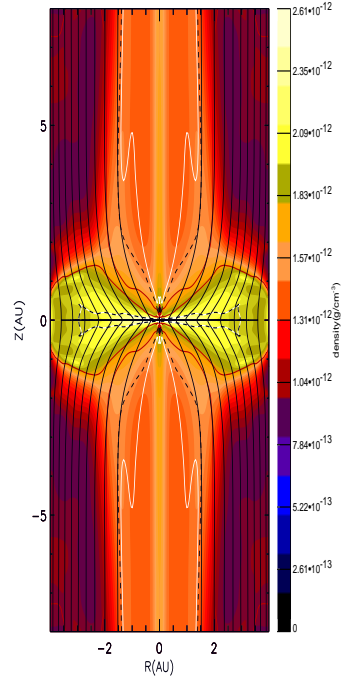}}}}
\end{center}
\caption{Same as in Fig.\ref{Fig1} but with a
  ideal stellar wind emitted from the inner region whose ejection mass
  rate is $\dot{M}=10^{-9}M_\odot/yr$. The disc-driven jet conserves a
  very similar dynamical structure to the case where no stellar wind 
  is emitted.
The size of the sink region
is $R_i=0.1 {\rm AU}$ and the stellar mass is $1 M_{\odot}$.}
\label{Fig3}
\end{figure}
The boundary conditions here are similar to CK04. We designed an absorbing sink
around the origin in order to avoid the gravitational singularity. In the first
quadrant of 
the simulation, the sink region is a square of one unit length both
in the $R$ and $Z$ directions where matter can only enter the zone
($V_R,V_Z\leq 1$) in order to avoid 
any numerical artifact. We consider the axis and disc mid-plane as a
combination of symmetric and antisymmetric boundaries. The top and right
boundaries are set as free boundaries (with nil gradients) except for the
outer radius of the disc where we impose a fixed poloidal mass accretion
rate.\\
The numerical simulations presented in this paper were performed using the
Versatile Advection Code VAC \citep{Toth96}, see
{\tt http://www.phys.uu.nl/$\sim$toth}. We solve the full set of
resistive and 
viscous MHD equations under the assumption of  cylindrical symmetry. We
time-advance the initial conditions using the conservative, second-order
accurate total variation diminishing Lax-Friedrichs scheme
\citep{TothetOdstreil96} with minmod limiting applied to the primitive
variables. We apply a projection scheme prior to each time step to enforce
 $\nabla\cdot\vec{B}=0$ \citep{BB80}. 
\subsection{Angular momentum transport in resistive, viscous thin accretion
  discs launching MHD jets}
\begin{figure*}[t]
\begin{center}
{\rotatebox{0}{\resizebox{4.cm}{9cm}{\includegraphics{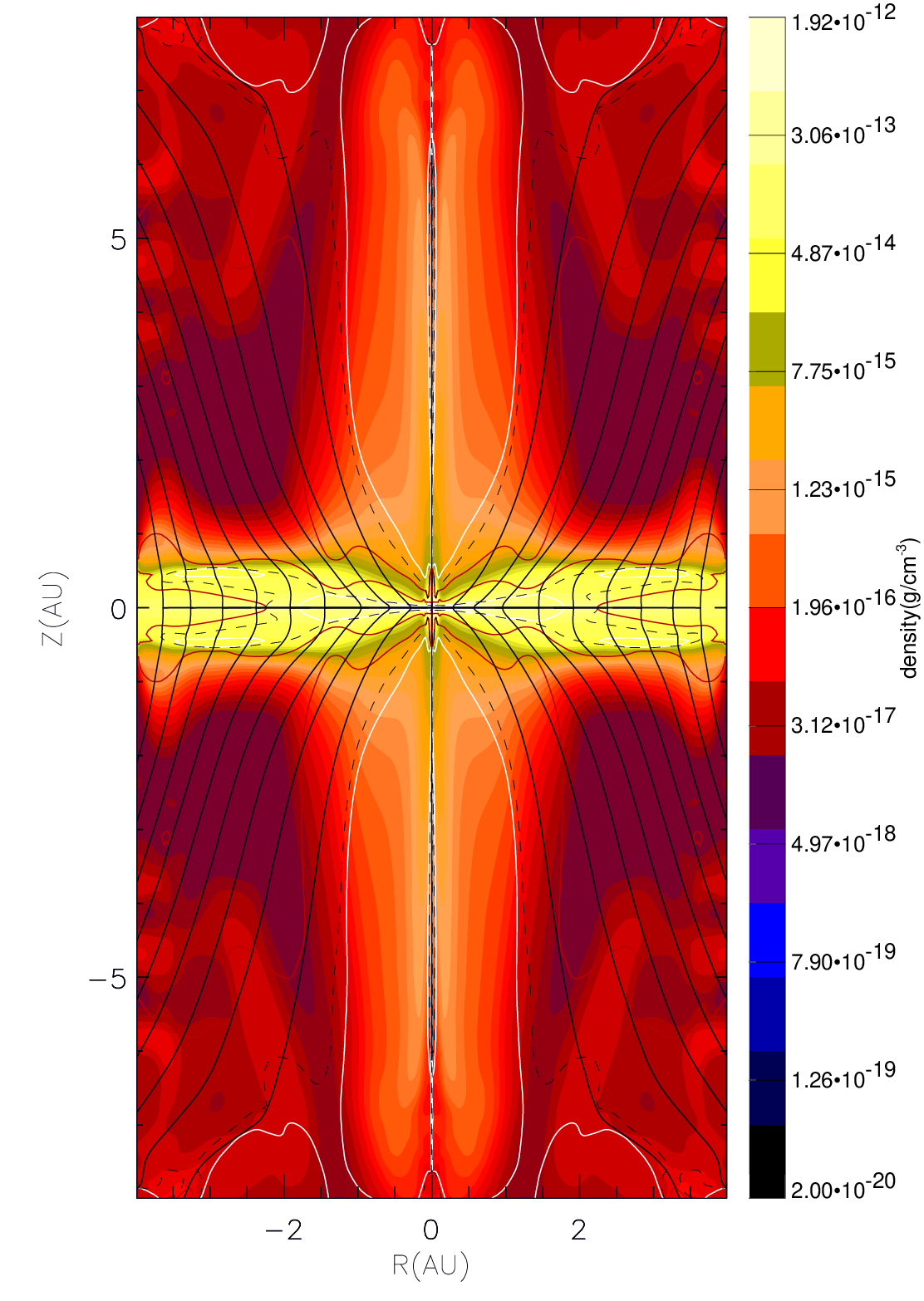}}}}
{\rotatebox{0}{\resizebox{4.cm}{9cm}{\includegraphics{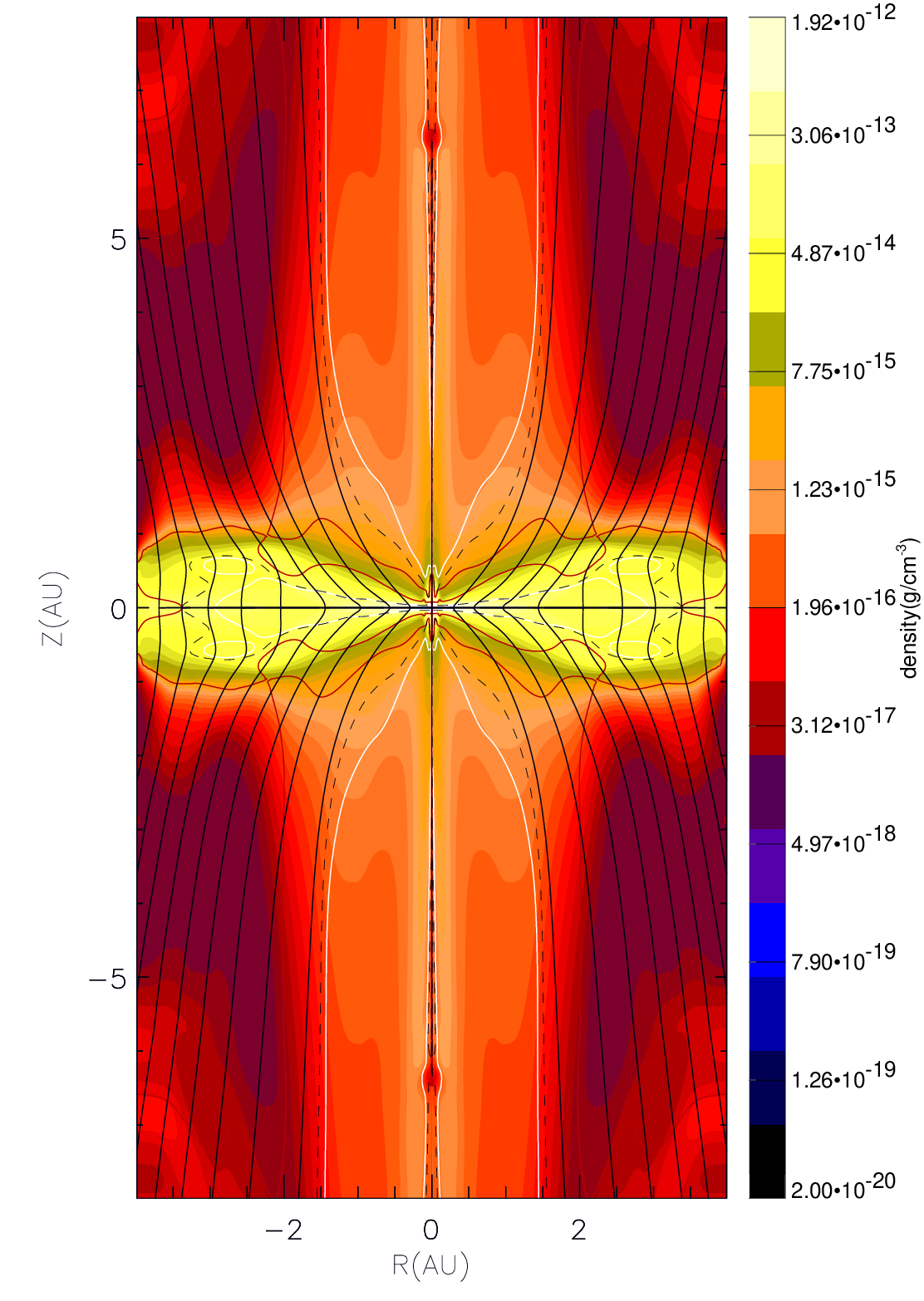}}}}
{\rotatebox{0}{\resizebox{4.cm}{9cm}{\includegraphics{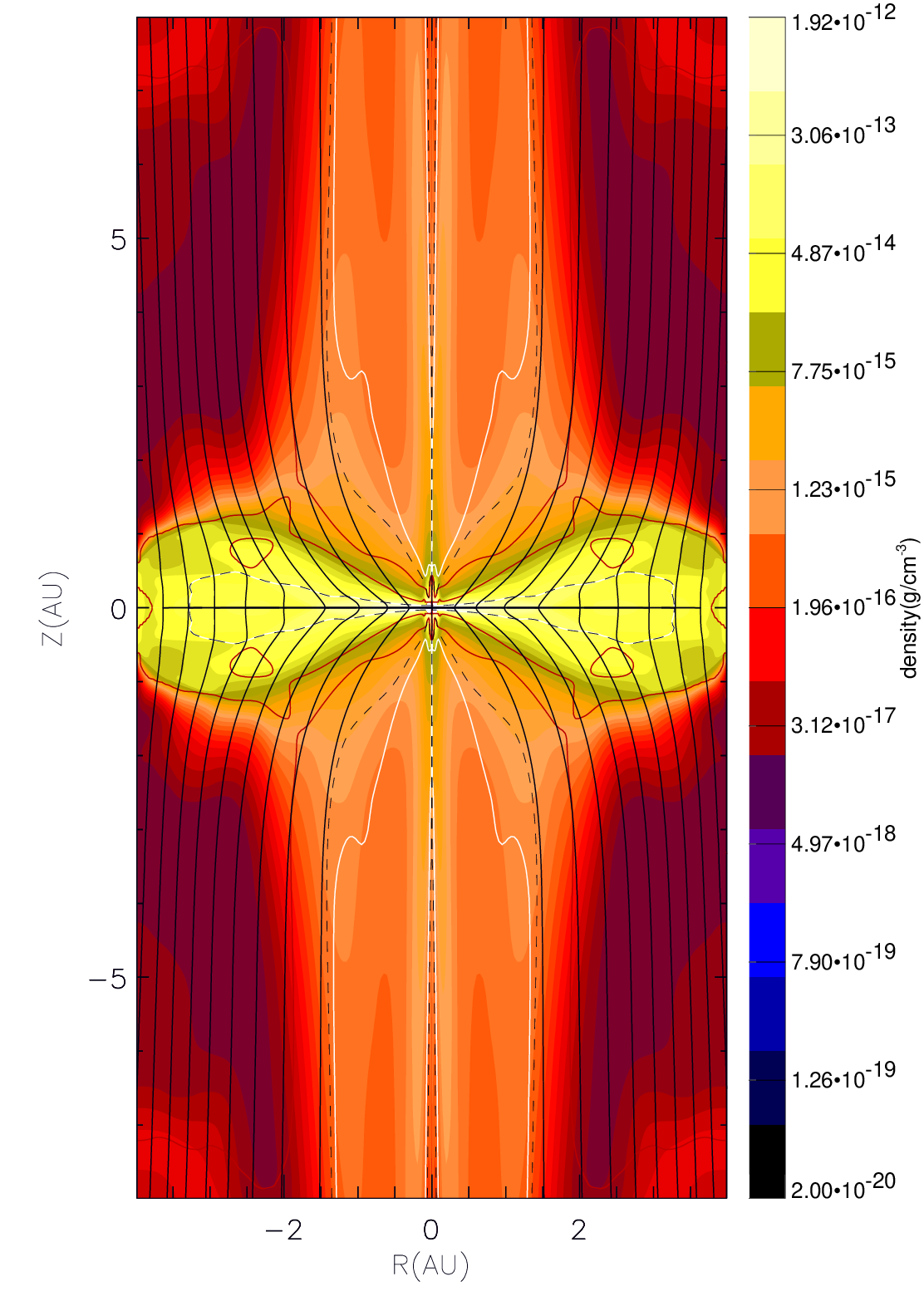}}}}
{\rotatebox{0}{\resizebox{4.cm}{9cm}{\includegraphics{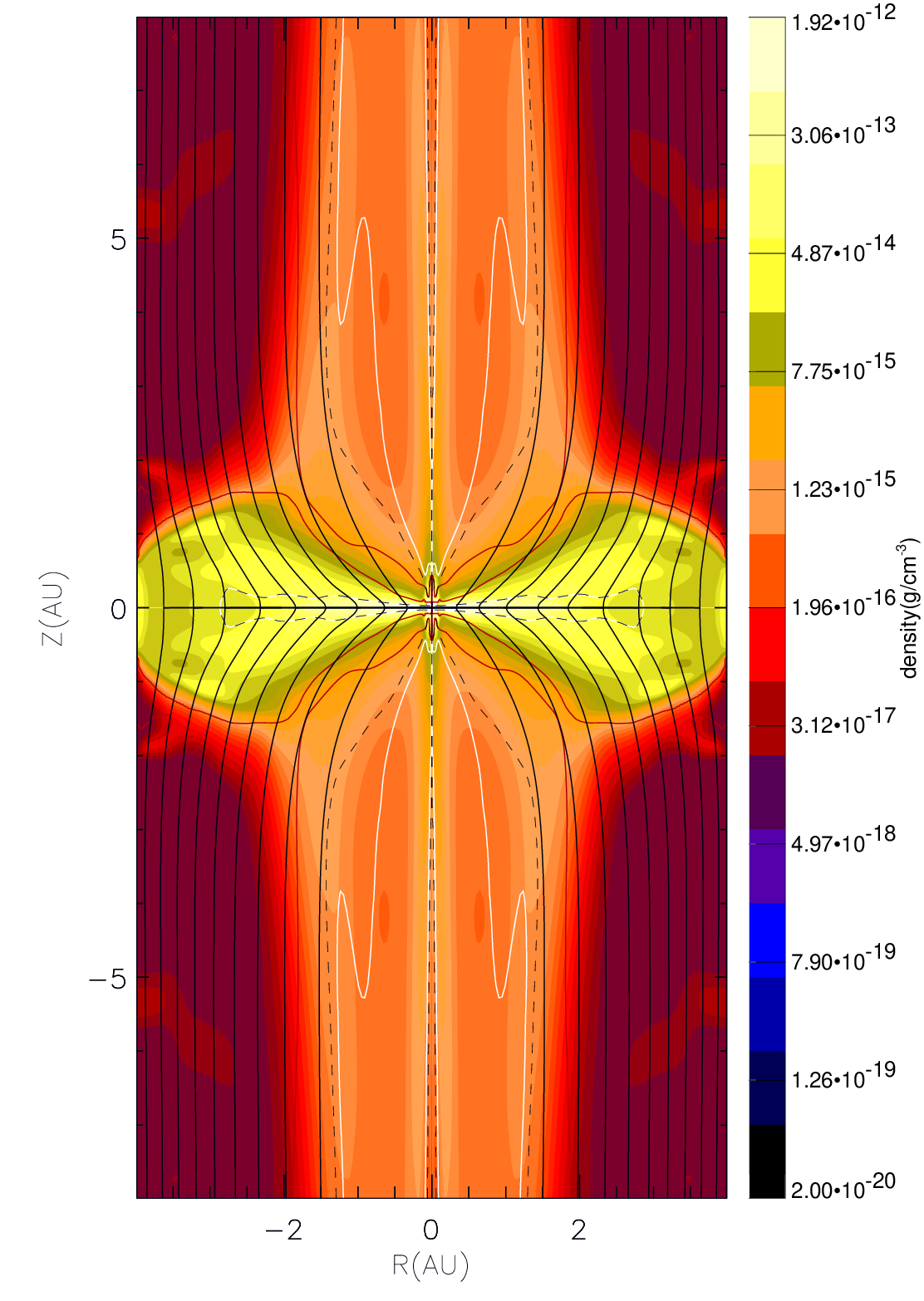}}}}
\end{center}
\caption{Time evolution of a two-component jet launched from a thin accretion
  disc threaded by a bipolar 
   magnetic field. The outflow is composed of a disc-driven jet embedding a non-ideal stellar wind 
   emitted from a YSO located at the center of the
   simulation in the sink region. The density
   contours are represented by greyscales while poloidal magnetic
   field lines are displayed using solid lines. The various snapshots
   represent the same system but at different stages (after five, ten, twenty,
   and thirty inner disc rotations).The simulation was performed with a
   stellar mass-loss rate of $\dot M=10^{-9}M_\odot/yr$. In the early stages of
   our simulation the stellar wind is the only outflow of the system while
   as the simulation goes on, a disc-driven jet appears around the
   stellar outflow and collimates it.
   The size of the sink region
   is $R_i=0.1 {\rm AU}$ and the stellar mass is $1 M_{\odot}$.}
\label{Fig4}
\end{figure*}
In order to study the angular momentum transport governing such an accretion
disc, we direct our first simulation here to studying the sole resistive and
viscous accretion disc threaded by a large-scale magnetic field.  We do not 
set any additional outflow coming from the central star in this simulation.\\
Writing down the conservation of angular momentum in an axisymmetric
framework, we see the two mechanisms responsible for the angular momentum
transport and removal, namely the magnetic torque and the viscous torque
\begin{equation}
\frac{\partial \rho V_{\theta}}{\partial t} + \nabla\cdot\left(\rho RV_{\theta}\vec{V}_{p} - 
RB_{\theta}\vec{B}_{p} + R\rho\eta_v\vec{\Pi}_{p\theta} \right) = 0
\end{equation}
where the subscript "$p$" stands for the poloidal component of the labeled
vectors. These two toroidal forces allowfor angular momentum transport in 
two different directions. Indeed, the magnetic torque provides angular 
momentum along the poloidal magnetic field, namely in the jet direction.
Since thin accretion-disc plasma velocity is dominated by the Keplerian
rotation, the viscous stress tensor can be approximated as 
\citep{Shakura&Sunyaev73}
\begin{equation}
R\eta_v\rho\vec{\Pi}_{p\theta} \simeq
\eta_vR²\rho\frac{\partial\Omega_K}{\partial R}\vec{e}_R =
-\frac{3}{2}\rho\eta_vR\Omega_K\vec{e}_R   
\end{equation}
where $\Omega_K$ is the Keplerian angular velocity. The viscosity will then
initiate a radial angular momentum transport. The range of turbulence
configuration in the accretion disc is quite endless (as the range of
Prandtl number), so we restrict ourselves to the configuration predicted in
the case of fully developed turbulence \citep{Pouquetetal76,Kitchatinov&Pipin94}.\\   
The result of the simulation is displayed on
Fig.\ref{Fig1} 
logarithmic density;
the poloidal magnetic field lines. The
obtained accretion-ejection structure shows a super-fastmagnetosonic
collimated jet. In order to study the difference with the  accretion-ejection
flows obtained by CK04, where the magnetic Prandtl number was set to zero, 
we measured the various angular
momentum fluxes crossing the internal and external radii, as well as through
the disc surface. To do so, we define global variables characterizing the
angular momentum extracted from the accretion disc, namely, 
\begin{equation}
L_{LIB} = L_{MEC} + L_{MHD} + L_{VIS}
\end{equation}
where
\begin{equation}
L_{MEC} = - \iint_{S_I} \vec{{\rm d} S_{I}}\cdot\rho\vec{v} R V_{\theta}
- \iint_{S_{E}} \vec{{\rm d} S_{E}}\cdot\rho \vec{v} R V_{\theta}
\end{equation}
is the variation of the angular momentum advected by the inflow between the
external and the internal parts of disc, 
 \begin{equation}
L_{MHD} = - \iint_{S_{I}}\cdot \vec{{\rm d} S_{I}}\vec{B}_p R B_{\theta}
- \iint_{S_{E}} \vec{{\rm d} S_{E}}\cdot\vec{B}_p RB_{\theta}
\end{equation}
is the variation of the MHD Poynting flux between the internal and external
radii. This magnetic contribution to the radial angular momentum accounts
for the twisting of the magnetic field occurring inside the disc, which is
similar to storing angular momentum and mechanical  energy of the plasma in
the magnetic field (generating toroidal magnetic field $B_{\theta}$).  
The amount of disc angular momentum removed by viscosity is given by 
 \begin{equation}
L_{VIS} = 
- \iint_{S_{I}} \vec{{\rm d} S_{I}}\cdot \vec{e_{R}}\rho\eta_{v} R \Pi_{R\theta}
- \iint_{S_{E}} \vec{{\rm d} S_{E}}\cdot \vec{e_{R}}\rho\eta_{v} R \Pi_{R\theta}.\,
\end{equation}
We denoted by  $\vec{{\rm d} S_{E}} = 2  \pi R_{E} {\rm d}Z \vec{e}_{R}$ the
outer  and by  $\vec{{\rm d} S_{I}} = 2  \pi R_{I} {\rm d}Z \vec{e}_{R}$
the inner  vertical cut through the accretion disc, with $-H <Z< H$.\\
When the outflow is arising from the accretion disc, we can evaluate the
angular momentum  transported vertically into the jet by considering the
various fluxes through the disc surface, 
\begin{equation}
L_{JET} = L_{MEC,J} + L_{MHD,J} + L_{VIS,J}
\,,
\end{equation}
where
\begin{equation}
L_{MEC,J} =  \iint_{S_{surf}} \vec{{\rm d} S_{surf}}\cdot\rho \vec{V}_p r V_{\theta}
\end{equation}
is the angular momentum advected by the vertical mass flowing into the jet,
 \begin{equation}
L_{MHD,J} =  \iint_{S_{surf}} \vec{{\rm d} S_{surf}}\cdot\vec{B}_p R B_{\theta}
\end{equation}
is the angular momentum extracted by the magnetic torque in the accretion
disc and converted into MHD Poynting flux through the disc surface. Finally,
 \begin{equation}
L_{VIS,J} =  \iint_{S_{surf}} \vec{{\rm d} S_{surf}}\cdot
\vec{e_{Z}}\rho\eta_{v} R \Pi_{z\theta}
\end{equation}
is the angular momentum extracted by the viscous torque at the disc
surface where $\eta_{v} \Pi_{z\theta} =- \eta_{v}
R\frac{\partial \Omega}{\partial z}$.   However, the effect of this
mechanism is zero, as  viscosity vanishes outside the accretion disc
(Eq. \ref{Eq_Resistivity_Disc}).  We note  the disc surface as $ \vec{{\rm
d}S_{surf}} = 2 \pi R {\rm d}R \vec{e}_{Z}$ with  $R_{I}<R<R_{E}$. In the
case where the whole structure reaches a stationary state, the angular
momentum conservation can be translated into a global relation
$L_{LIB}=L_{JET}$.   The angular momentum fluxes related to our simulation
are displayed in Fig.\ref{Fig2}, where we represent $L_{MHD}$, $L_{VIS}$ 
and $L_{JET}$ normalized to the flux of angular momentum removed from the
disc $L_{MEC}$. It clearly shows that inside a resistive, viscous, thin 
accretion disc with Prandtl number equal to unity, the viscosity
 is unable to remove the disc angular
momentum efficiently  (as already showed by Pudritz \& Norman 1986), since only one
percent is carried away by the viscous torque. 
Conversely the presence of the jet has an strong impact on the angular
momentum balance because it enables the magnetic torque to achieve a very
efficient angular momentum transport from the disc into the jet (more than
$90\%$). Among the three components of $L_{JET}$, $L_{VIS,J}$ is nul
because viscosity is vanishing at the disc surface and $L_{MEC,J}$ is small
compared to the MHD Poynting jet flux $L_{MHD,J}$ (mainly because mass is
sub-slow-magnetosonic at the disc surface). This magnetic energy reservoir
created at the base of the jet is used in the jet to accelerate matter such
that the jet becomes super-fast-magnetosonic. It is noteworthy that the
structure coming from our simulation reaches a quasi-stationary state where
$L_{JET}\simeq L_{LIB}$.\\

\begin{figure*}[t]
\begin{center}
{\rotatebox{0}{\resizebox{6cm}{11cm}{\includegraphics{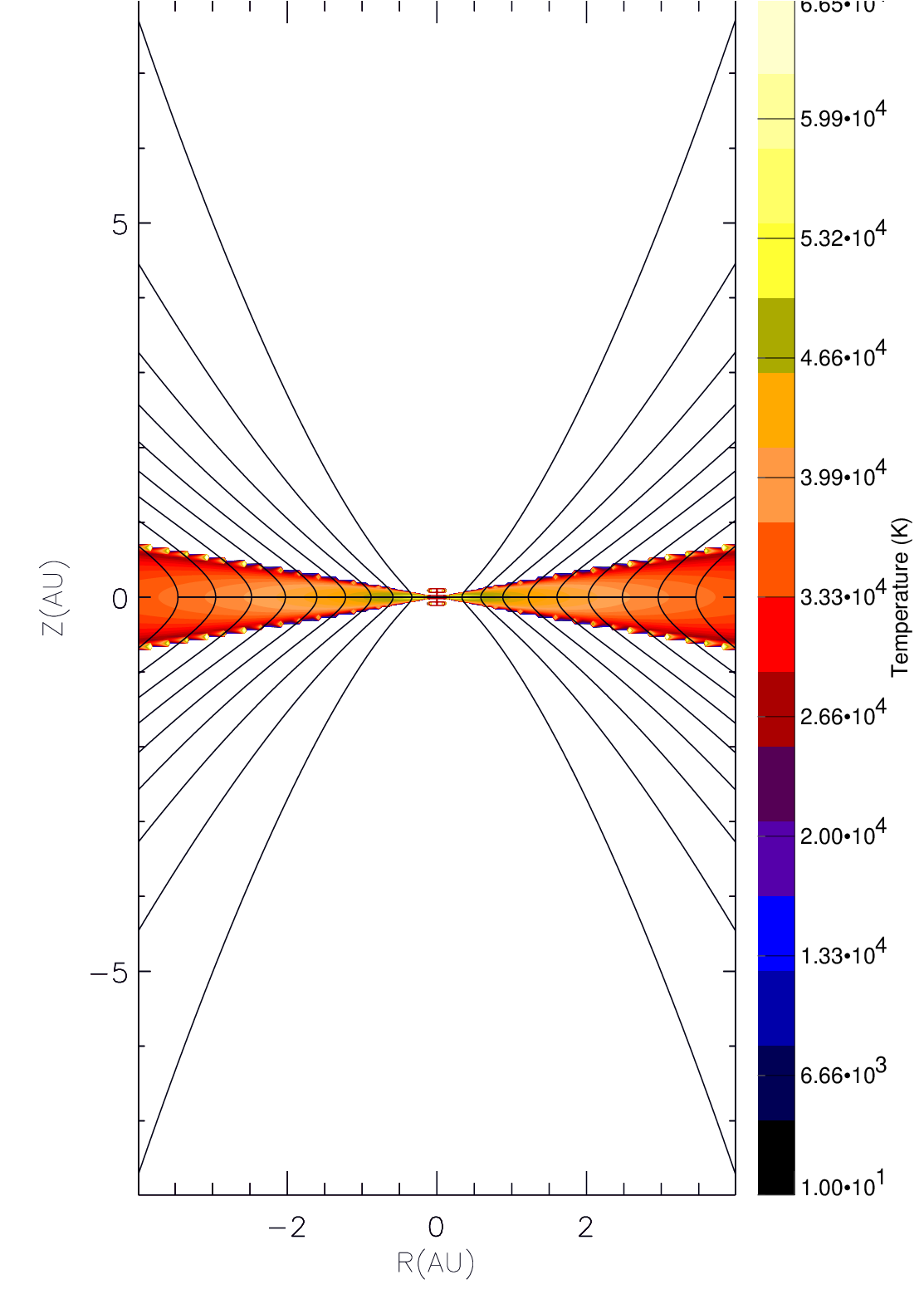}}}}
{\rotatebox{0}{\resizebox{6cm}{11cm}{\includegraphics{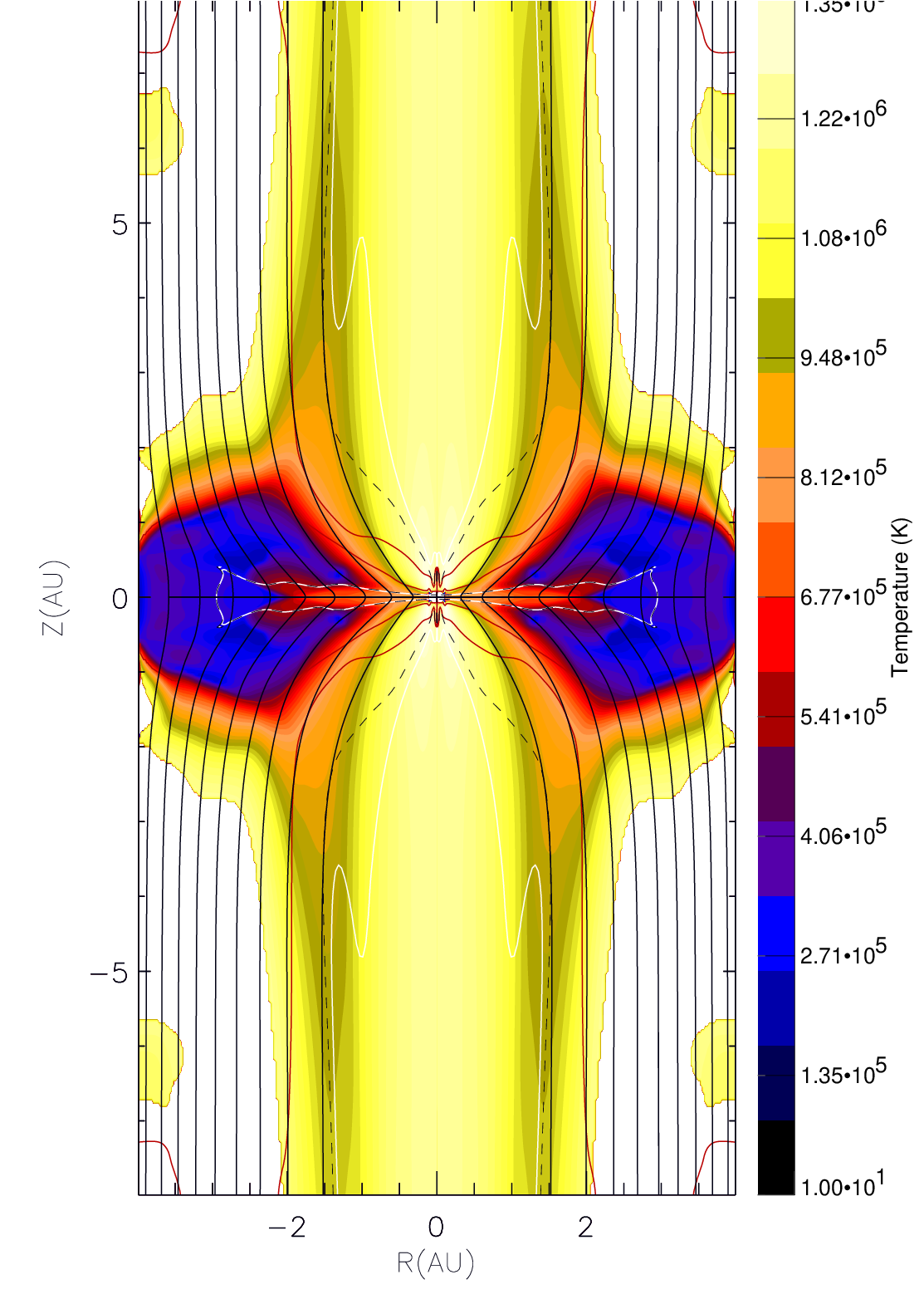}}}}
\caption{ {\bf Left:} plot of the initial temperature isocontours of the
  accretion disc.  
 {\bf Right:} plot of the  temperature isocontours of the simulation displayed in
  Fig.\ref{Fig3}. Temperature isocontours are not displayed in the
  external medium (outside of both the jet, stellar wind, and disc) as it is
  considered  a near vacuum medium with very low temperature.
The size of the sink region
is $R_i=0.1 {\rm AU}$ and the stellar mass is $1 M_{\odot}$.}
\label{Fig5}
\end{center}
\end{figure*}

In order to explain why the magnetic torque prevails in the accretion disc,
we can use some ordering to estimate the relative amplitude of the two
torques. The magnetic torque expression is 
\begin{equation}
(\vec{J}\times\vec{B})\cdot\vec{e}_{\theta}=B_Z\frac{\partial
 B_{\theta}}{\partial Z} + \frac{B_R}{R}\frac{\partial
 RB_{\theta}}{\partial R}
\end{equation}
that we can simplify knowing that $B_{R}\ll B_Z$ inside the disc and that
the toroidal magnetic field that is nil at the disc mid-plane approaches $-B_Z$
on the disc surface,
\begin{equation}
|(\vec{J}\times\vec{B})\cdot\vec{e}_{\theta}|\simeq \frac{B^2_Z}{H} \ .
\end{equation}
The viscous torque, as previously written, can be expressed as
\begin{equation}
\left|\nabla\cdot (R\eta_v\vec{\Pi}_{p\theta}\cdot\vec{e}_{\theta})\right|
\simeq 
\left|\frac{\partial \rho\eta_vR\Omega_K}{\partial R}\right| \simeq
\frac{\Omega_K\rho\eta_v}{R}
\end{equation}
since the radial variation of the various quantities is expected to be like
power laws \citep{Shakura&Sunyaev73}. The ratio of the two torques can be
written as 
\begin{equation}
\left|\frac{(\vec{J}\times\vec{B})\cdot\vec{e}_{\theta}}{(\nabla\cdot
  R\eta_v\vec{\Pi}_{p\theta})\cdot\vec{e}_{\theta}}\right| \sim
  \left(\frac{V_A}{C_S}\right)^{2}\frac{1}{\alpha_v\epsilon}  
\end{equation}
where the disc aspect ratio $\epsilon$ is much lower than unity in a thin
accretion disc, as is the viscosity parameter $\alpha_v$. The accretion
disc has to be close to equipartition between magnetic pressure and
thermal pressure in order to launch a jet \citep{Ferreira&Pelletier95}, so
it is easy to see that this ratio is much higher than unity in all
magneto-viscous, thin-disc launching jets. In the context of our
simulation, we have set $\epsilon=\alpha_v=0.1$ leading to a ratio on the
order of $10^{2}$, which is compatible with the ratio of $L_{JET}$ to
$L_{VIS}$ in Fig.\ref{Fig2}.    
In conclusion, we see that angular momentum extraction from a thin or even a 
slim magnetized-disc ($\epsilon \ll 1$) is likely to occur 
in the disc-driven jet rather than in the disc itself, for a disc close 
to equipartition, i.e. with a plasma beta close to one.
This results is consistent with 
previous analytical models of non-resistive disc winds  where the
accretion-related 
wind removes the excess of the angular momentum 
\citep{Pudr86, Pelletier&Pudritz92, Lubowetal94}. \\
 However, in order to 
have a more consistent simulation of the accretion-ejection structures, 
we should take the interaction with the inner stellar coronal wind into account.
Incontrast to previous simulations, we include the acceleration of the 
stellar wind, which  
is likely to start with a subsonic motion from the base of the corona and then
accelerates, as well as the full description of the accretion-disc launching 
jets. The stellar wind  acceleration close to the axis cannot be exclusively 
magnetic, since magneto-centrifugal effects vanish near the axis. Besides, the
high coronal temperature is
likely to induce a more efficient turbulent heating. We thus intend to use the
turbulent wind viscosity and resistivity as the primary sources of
acceleration for the inner stellar outflow.
Turbulence may be induced in the stellar magnetospheric wind  by its interaction
with the disc-driven jet. The differences between both their dynamics
and thermodynamics probably induce instabilities.
The turbulence may also have a stellar origin and/or a possible connection
to the accretion occurring near the surface of the star. In fact, the inner
accretion surface, as well as the star surface, are time-dependent and
inhomogeneous, leading to outwardly propagating  Alfv\'en
waves in the stellar wind and inducing turbulence.
This origin of turbulence is based on an analogy with  models and
observations of the solar wind where the solar origin of turbulence is
investigated  \citep{Leamonetal98, Smith&Matthaeus&Zank00} as the convection
below the photosphere (Cranmer \& Ballegooijen 2005;
 (See also a review papers \citep{Goldsteinetal95, Cranmer04}). 
\section{Two-component MHD outflows from a resistive, viscous, accretion 
disc and a star corona}

\noindent 
The aim of this section is to  model the interplay between the two components 
of a YSO outflow, namely a jet launched from a magnetized resistive, viscous
accretion disc and the second one, an non-ideal MHD spherical wind ejected 
from the protostar hot corona. In this section, we describe the method that 
we developed for the implementation of the stellar wind in the
model. 
The results and the differences, with the non-ideal MHD stellar wind model,
is discussed in the following section. 
We use the same initial conditions as in the previous section, except for 
is a change in the boundary condition located at the top of the sink region. 

\begin{figure}
\begin{center}
{\rotatebox{0}{\resizebox{8cm}{6cm}{\includegraphics{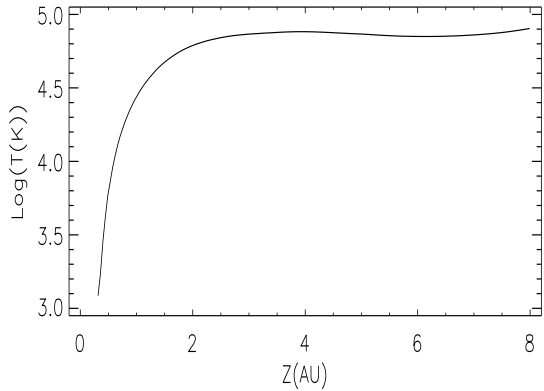}}}}
\caption{Plot of the vertical variation  at $R=1{\rm AU}$ 
from the axis of  the  temperature in the simulation displayed in
Fig.\ref{Fig5}. 
 The size of the sink region
is $R_i=0.1 {\rm AU}$ and the stellar mass is $1 M_{\odot}$.} 
\label{Fig6}
\end{center}
\end{figure}
\begin{figure*}[t]
{\rotatebox{0}{\resizebox{18cm}{6.5cm}{\includegraphics{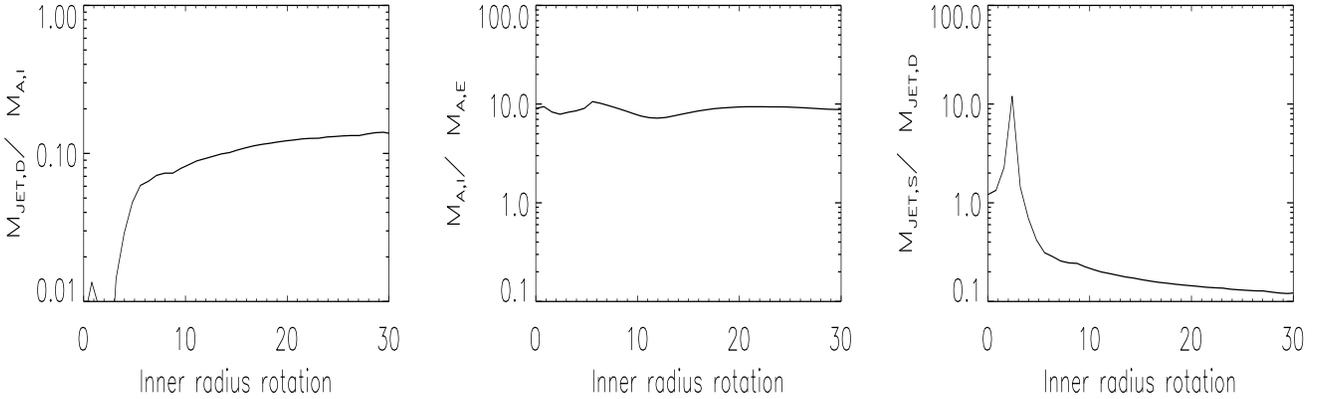}}}}
\caption{{\bf a)} Plot of the ejection mass  rate from the accretion disc.
  {\bf b)} the inner 
  accretion rate. {\bf c)} the ratio of the stellar mass-loss rate to the
  disc-wind mass-loss rate 
 as a function of time. These plots are related to the simulation performed
 with a  stellar mass loss  of
$\dot{M}=10^{-9}M_\odot/yr$.}
\label{Fig7}
\end{figure*}

We now replace the accretion inflow by an outward mass flux whose amplitude is
$\tau$ times the solar mass-loss rate and which is spherically ejected with a 
speed that is a fraction $\delta$ of the fast-magnetosonic speed in the corona
$V_{f}^{in}$;
\begin{eqnarray}
\dot{M} &=& \tau 10^{-14} M_{\odot}/yr\,
\label{Limite_Masslossup}\\
v_R^{up}  &=& \delta_{v}\, V_{f}^{in}  \frac{R}{Z} \,
\label{Limite_Vrup}\\
v_Z^{up}  &=& \delta_{v}\, V_{f}^{in}\,.
\label{Limite_Vzup}
\end{eqnarray}
 In our computational domain, the sink region around the origin is a square of one 
unit length in both the $R$ and $Z$ directions. This length corresponds to the
internal radius $R_i$ of the accretion disc but does not correspond to the
physical radial disc edge that is located inside the sink. In the
following plots, we set $R_i$ to $0.1 AU$ so that the upper limit
of the sink region and our internal disk radius are at $20$ star 
radii from the star surface in the case of a one solar-mass central object.
The magnetic field near the star has a near spherical  expansion that is
becoming a near vertical structure (Eq.\ref{Magn_prescription}). Our
magnetic field prescription is coherent with a magnetic field at the
surface of the star on the order of  $B\sim 2 {\rm kG}$ in agreement with
observational values \citep{Johns-Krulletal01, Guenttheretal99}. 

 In our simulations, we model neither the very inner part of the
 accretion disc  nor its interaction with the magnetosphere of the star
 located at ($3\,R_{\odot}-6\,R_{\odot}$).

Nevertheless, we include the effect of the star rotation in our simulations
by imposing a  solid rotation profile ($V_{\theta}/R=cst$) on the outflow at
the top of the sink region. We set the angular velocity of the outflow at 
this boundary to the Keplerian angular velocity at the inner radius $R_i$.
The rotation period associated is then
\begin{equation}
P_{rot} = 11.57\, days \left(\frac{R_i}{0.1AU}\right)^{3/2}
\left(\frac{M_*}{M_{\odot}}\right)^{-1/2}\,,
\end{equation}
which corresponds to young star rotation periods such as  GM Tau
\citep{Gullburingetal1998}. 

We also consider two different cases of stellar winds. The
first one is consistent with a heavy  hot wind whose mass loss is on the
order of  $\dot{M} = 10^{-7} M_{\odot}/yr$ ($\tau=10^{7}$). This mass-ejection
rate is in the range of typical mass losses for young B-star and O-star type. 
This kind of YSO is characterized by strong outflows and dynamical timescales 
around $10^{4} years$, and the stellar wind is believed to be the main 
contributor to the outflow. The computational domain related to this simulation
is $\left[R,Z\right] = \left[0,80\right] \times \left[0,120\right]$ 
with a resolution of $304 \times 404$ cells.
The other simulation stands for systems where the stellar wind is a light
and hot one, namely with a mass
loss on the order of $\dot{M} = 10^{-9} M_{\odot}/yr$ ($\tau=10^{5}$) 
(Fig.\ref{Fig3}). 
In  these cases  the stellar wind is lighter than the jet launched from
the accretion  disc. The computational domain associated with the second
simulation is $\left[R,Z\right] = \left[0,40\right] \times
\left[0,80\right]$  
with a resolution of $134 \times 204$ cells. It is noteworthy that, since we
are expecting a widening of the jet due to the strong stellar mass loss in
the first simulation, we have designed a larger computational domain in order
to capture all the features of the resulting outflow. 
We set the velocity parameter $\delta_{v}$ to $0.01$ in both simulations, 
which is consistent with an initial sub fast-magnetosonic and sub-Alfv\'enic 
ejection from the corona.
We assume that part of the outflow acceleration has already taken place between
the star surface (hidden in the sink) and the top boundary of the sink region
although the flow is sometimes even sub-slow-magnetosonic, depending on the
magnetic configuration.
\subsection{Non-ideal effects in stellar winds}

In most stellar wind models, the wind material is often subject to coronal 
heating, contributing to the global acceleration of the flow.
 In our simulations, we assume that the coronal heating is a 
fraction $\delta_{\varepsilon}$ of the energy released  in the 
accretion disc at the boundary of the sink region, which is  
transformed into thermal energy in the stellar corona close to the polar axis. 
This scenario was proposed by \citep{Matt&Pudritz05} and is
supported by the current observations of hot stellar outflows \citep{Dupreeetal05}. 
The thermal energy imposed at the lower boundary of the corona  
is, at each step of the simulation, the sum of the
thermal energy $\varepsilon^{up+1}$ of the  above stellar jet, i.e. the thermal energy of the 
first cells above the sink border, plus a fraction $\delta_{\varepsilon}$
of  the thermal energy at the disc inner radius $\varepsilon^{in}$. 
Thus the thermal energy at the upper boundary of the sink is
\begin{eqnarray}
\varepsilon =   \varepsilon^{up+1}+   \varepsilon^{in} \delta_{\varepsilon}
\,.
\end{eqnarray}
The $\delta_{\varepsilon}$ parameter range is limited from below by the initial 
thermal acceleration  at the surface of the corona, which should balance the
gravitational force, and from above by the requirement of avoiding too high a 
temperature in the corona. 
In our simulation we take a small efficient 
heating corona $\delta_{\varepsilon}=10^{-5}$. 
We deliberately use a very low value for this parameter in order to ensure that 
the main heating source of the stellar wind lies in the Ohmic heating. We intend
  to study its effects compared to the prescription of a larger amount of
  thermal energy  at the base of the stellar flow. The low value of
  $\delta_{\varepsilon}$ represents the amount of energy released by
  Ohmic heating below the surface of the sink. This heating is essential
  for letting the flow escape from the gravity,
 since the flow has already undergone an initial acceleretion
from the surface of the star to the top of the sink region. Moreover, the wind undergoes a mechanical heating where the accreted 
flow at the top of the accretion disc compresses the inner wind and may sustain 
turbulence in the wind.   In Sect. 3.2.2, we will discuss the ideal stellar
winds emitted from the sink where a large amount of thermal energy is
deposited at the base of the flow.  \\
The interaction between the different components of the outflow may be 
responsible for energy dissipation inside the plasma, which is the outcome of
non-ideal MHD mechanisms occurring in the wind.  
In this paragraph, we show how these non-ideal MHD effects are taken into
account when prescribing a turbulent magnetic resistivity
taking place in the wind region, in addition to the disc resistivity
\begin{eqnarray}
\eta_{m} &=& \alpha_{m} \left.V_{A}\right\vert_{Z = 0} H \exp\left(-2
\frac{Z^2}{H^2}\right) \nonumber\\ &+& \alpha_{w} V_{A} H_{w}  \exp{\left[-2
\left(\frac{R}{H_{w}}\right)^2\right]} \ .
\end{eqnarray}
The first term accounts for the anomalous resistivity occurring in the
accretion disc. It vanishes outside the disc ($Z>H$).  The second
term corresponds to the description of an anomalous resistivity occurring in
the  outflow close to its polar axis. This term vanishes outside the stellar wind
($R>H_{w}$) where $H_{w}$ is the distance from the polar axis where the Alfv\`en
speed encounters  a minimum.
Hence, the dissipation effects are only located in the stellar wind
component and not in the disc wind, which is supposed to be less turbulent.
For the resistivity in the stellar wind, we take $ \alpha_{w}= 10^{-2}$, a
lower value than in the disc itself. 
\begin{figure*}[t]
{\rotatebox{0}{\resizebox{9cm}{8cm}{\includegraphics{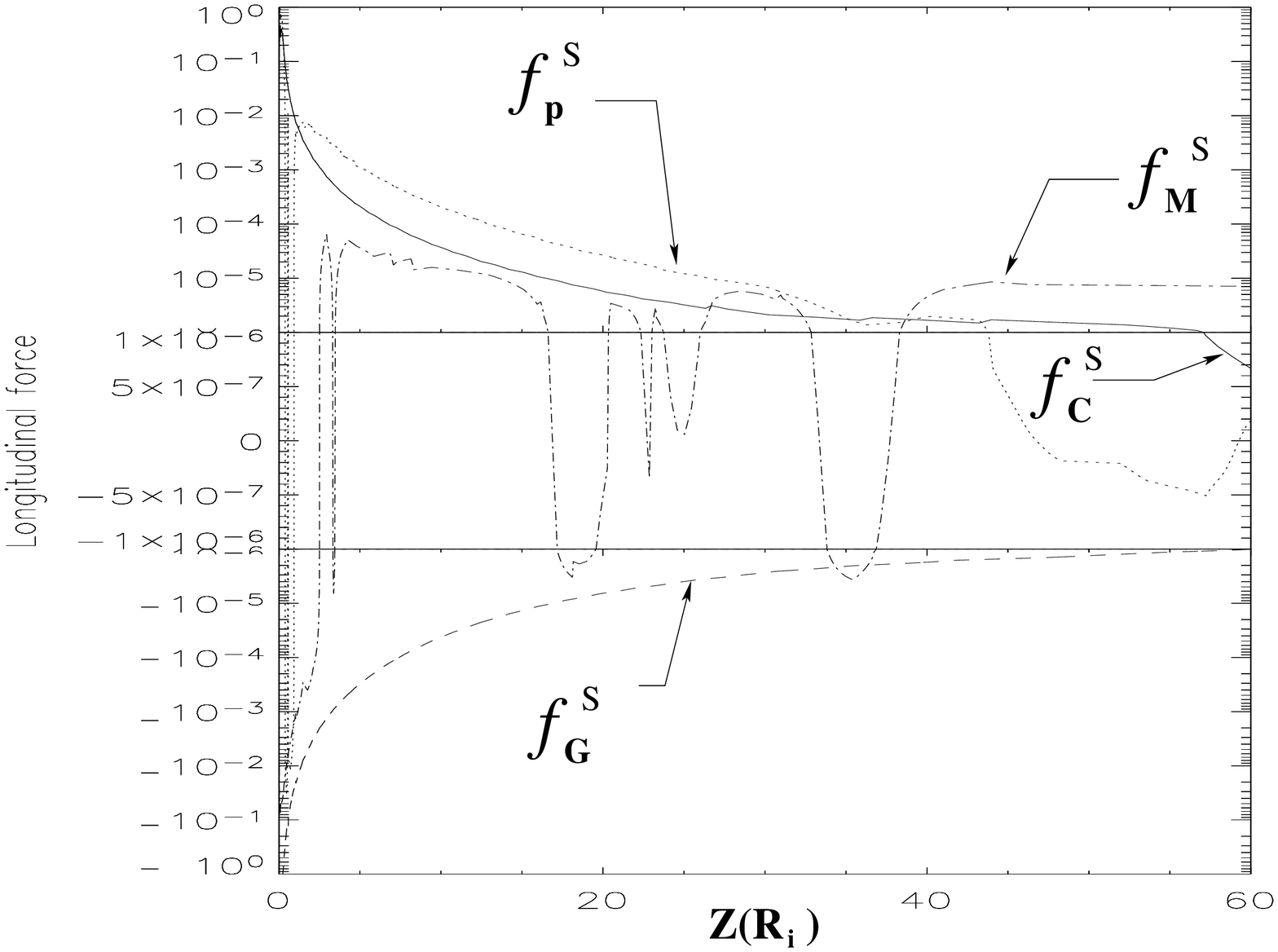}}}} 
{\rotatebox{0}{\resizebox{9cm}{8cm}{\includegraphics{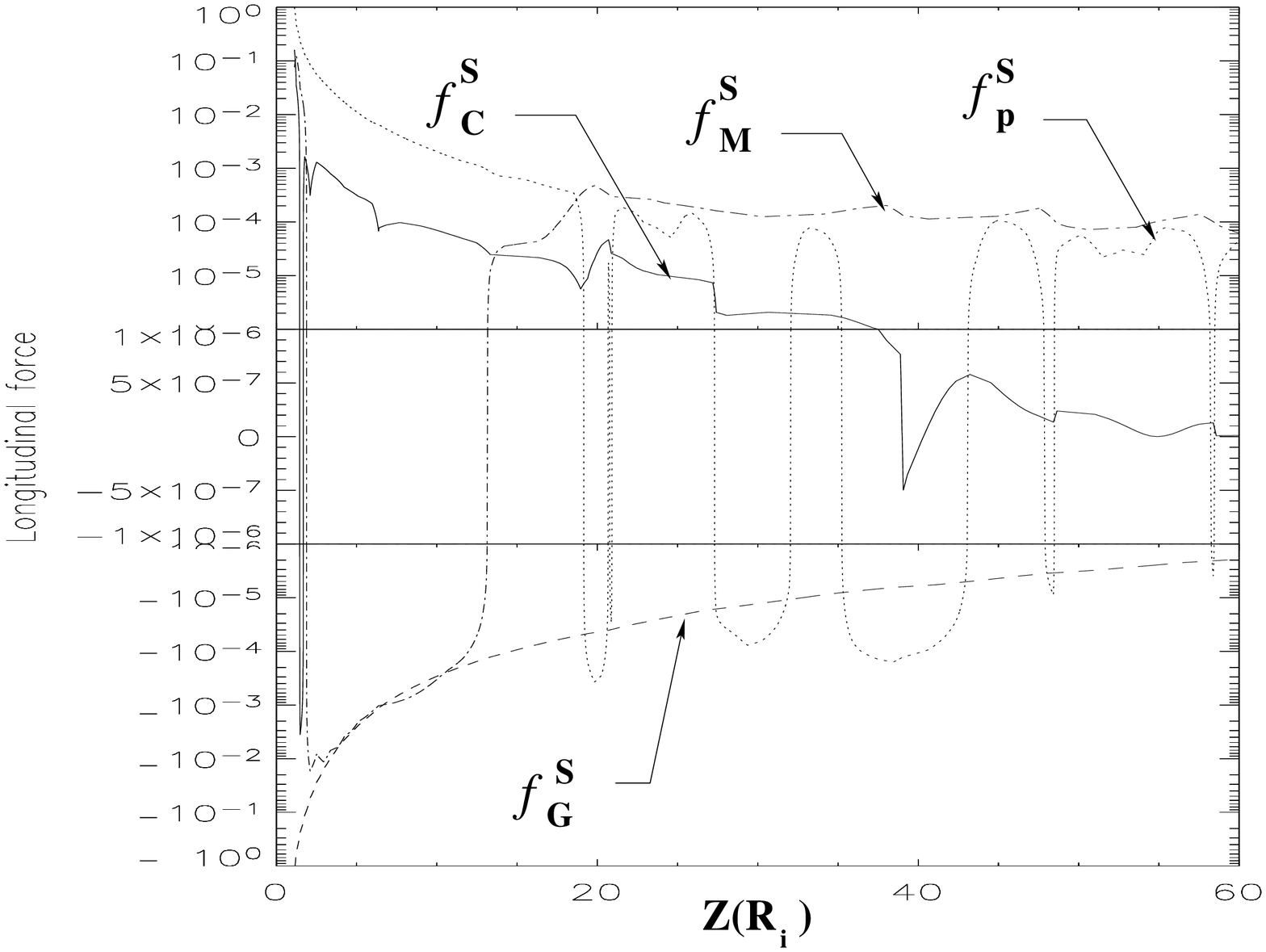}}}}
\caption{Plot of the various forces {\bf Left:} along a given magnetic field line 
 anchored in the accretion disc, {\bf Right:} along a streamline anchored
 to the stellar corona. These plots show the various forces accelerating the 
 flow:
 $f_C^s$ is the centrifugal force, $f_M^s$ the magnetic one, $f_P^s$ the
 pressure gradient, and $f_G^s$ the gravitational force.} 
\label{Fig8}
\end{figure*}
\subsection{Two-component MHD outflows from YSO}

We first focus on a simulation where the stellar mass loss is set to
$10^{-9}M_{\odot}/yr$. The outcome of our simulation can be seen in
Fig.~\ref{Fig4} where we display four different snapshots of the poloidal
cross-sections  of the structure
at different times  its evolution, i.e. at 8, 16, 24, and 32
rotation periods of the inner disc radius.  In these snapshots we
display the density  contours and the poloidal
magnetic field lines.  The initial accretion disc
configuration is close to a hydrostatic equilibrium where the centrifugal
force and the total pressure gradient balance
the gravity. In the central region, the matter is continuously emitted at
the surface of the sink region (designed to be close to the star surface)
with sub fast-magnetosonic speed and with a solid 
rotation  velocity profile.  Initially, a conical hot outflow (stellar wind)
propagates
above the inner part of the disc. Its inertia compresses the magnetic field
anchored to the accretion disc. As a result, the bending of the magnetic
surfaces increases, leading to a magnetic pinching of the disc. This pinching 
delays the jet launching as the disc has to find a new vertical equilibrium. 
Thus the disc takes a few more inner disc rotations before launching its jet
compared to CK04. 
Once the jet has been launched, the structure reaches a quasi steady-state 
where the outflow becomes parallel to the poloidal magnetic field that is 
parallel to the vertical direction.\\
The obtained solution is fully consistent with an accretion-disc launching  
plasma with a sub-slowmagnetosonic velocity. The solution crosses the three 
critical surfaces, namely the slow-magnetosonic, the Alfv\'en, and the 
fast-magnetosonic surfaces. The
other component of the outflow, namely the stellar wind, is injected with
sub-fastmagnetosonic velocity and crosses  the Alfv\'en and
fastmagnetosonic surfaces.  
The two components of the outflow become super-fastmagnetosonic  before
reaching the upper boundary limit of the computational
domain. Figure~\ref{Fig4} also shows that the outflow has achieved quite a 
good collimation within our computational domain. We can  distinguish between 
the two components using the isocontours of temperature, which are displayed 
as grey-scales in Fig.~\ref{Fig5}. In this figure, we can clearly see
a hot outflow coming from the central object embedded
in the cooler jet arising from the accretion disc.  In Fig.~\ref{Fig5}
  we also show that the thermal 
energy released  by the Ohmic and viscous heating in the accretion disc 
is extracted by a hot jet that is compatible with the result in
CK04. In order to illustrate the thermal effect on the outflow, we have plotted
the temperature vertical profile in Fig.~\ref{Fig6} along a radius located at 
$1 \,{\rm AU}$ from the axis. In this plot, the temperature increases
in the disc corona before  reaching its maximum $T=10^{5}{\rm K}$ and
  remaining constant. \\
To study the time evolution of both accretion and ejection
phenomena in the  accretion disc and around the star, we
analyze  the  accretion and ejection mass loss rate in both components.
As in CK04 we draw  the time evolution in Fig.~\ref{Fig7} of the mass loss 
rate $M_{jet,D}$ in the disc-driven jet normalized to the accreted mass rate 
$M_{A,I}$ at the inner radius  $R_I =1$. We also display  $M_{A,I}$ normalized
to the fixed mass accreted $M_{A,E}$ at the external radius of our
accretion disc at $R_I =40$. Similar to CK04, we observe a strong increase,
on the accretion rate in the inner part with time (Fig.~\ref{Fig7}). 
This behavior is related to the extraction of  the rotational energy of the 
accretion disc by the magnetic field. Indeed the creation of the toroidal 
component of the magnetic field in the disc brakes the disc matter so that 
the centrifugal force decreases, leading to an enhanced accretion motion. 
The mass flux associated with the
 disc-driven jet slowly increases to reach $18\%$
of the accreted mass rate at the inner radius and contributes to $98\% $ of the
total mass-loss rate of the outflow. In fact, in this simulation the mass-loss
rate from the central object is constant ($10^{-9}M_{\odot}/yr$), while the
inner accretion rate reaches 
$10^{-6}M_{\odot}/yr$ and the disc-driven jet mass rate
$10^{-7}M_{\odot}/yr$. Hence the stellar outflow does not affect the
overall structure of the outflow much. This is confirmed by the shape of the
outflow since it reaches a very similar aspect to the one obtain  in CK04 or
in the previous simulation without a stellar jet, i.e. a jet
confined within $20$ inner disc radius.\\
In order to analyze  this accretion-ejection engine, we have to identify the
forces responsible for the establishment of a steady  accretion motion in
equilibrium with a continuous emission of matter at the surface of the
accretion disc. Furthermore we have to look at the collimation of the
outflow and its interaction with the stellar wind.  We show in
Fig.~\ref{Fig8} the various forces parallel to a given magnetic field line 
anchored in the disc and a flow streamline anchored to the central object. The various
forces working along and across the field lines are 
\begin{figure}[t]
{\rotatebox{0}{\resizebox{9cm}{8cm}{\includegraphics{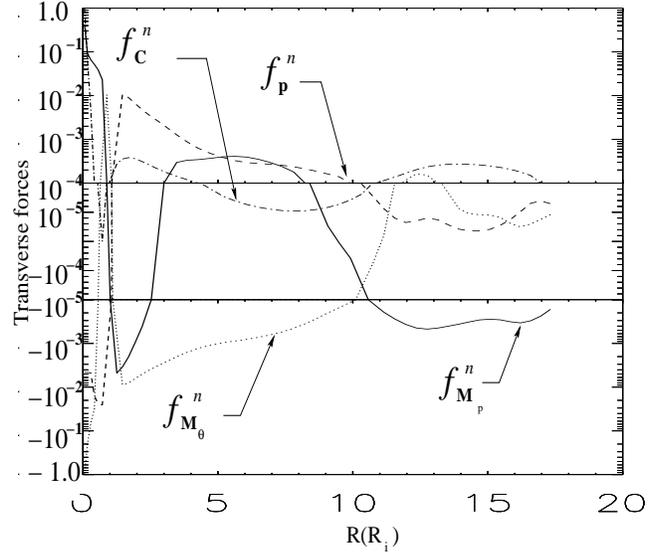}}}}
\caption{Plot of the transverse forces as a function of $R$ at a given
  altitude Z = 75. It shows, across a 
given cross section of the jet, the collimation processes acting in the
stellar and disc components 
of the jet for the simulation with a stellar mass
 loss  $M_\odot=10^{-9}M_\odot/yr$.  $f_C^n$ is the centrifugal force,
 $f_P^n$ the pressure gradient and 
$f_G^n$ the gravitational force.  $f_{M_\theta}^n$ the Lorentz force due to
  the toroidal component 
of the magnetic field, while  $f_{M_{\rm P}}^n$ corresponds to the poloidal component.}
\label{Fig9}
\end{figure}

\begin{figure}
{\rotatebox{0}{\resizebox{9.0cm}{8cm}{\includegraphics{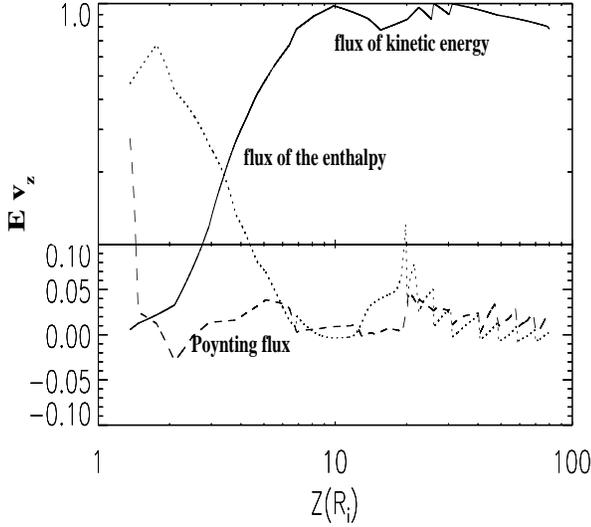}}}}
\caption{ The  vertical variations of specific energies, normalized to the
  maximum kinetic energy flux, along a streamline in the stellar 
wind with a rate $\dot{M}=10^{-9}M_\odot/yr$. This plot clearly illustrates
the thermal acceleration provided by the enthalpy to the stellar wind
material. The enthalpy gradient is here sustained by local turbulent Ohmic
heating representing, in our simulation, $35\%$ of the energy released by
accretion (see Sect.~\ref{Stellarwind})} 
\label{Fig10}
\end{figure}

\begin{eqnarray}
f_{\rm P}^{s} &=& - \frac{\partial P}{\partial s},\\ 
f_{\rm G}^{s} &=& -
\frac{G M}{(R^2+Z^2)^{3/2}}(R\vec{e_R}+Z\vec{e_Z})\cdot \vec{e}_{\rm s},\\
 f_{\rm C}^{s} &=&
V_{\theta} R \vec{e_R}\cdot \vec{e}_{\rm s},\\
  f_{\rm M}^{s} &=& - \frac{1}{2}
\frac{\partial (R B_{\theta})}{\partial s}. \\ 
f_{\rm P}^{n} &=& -
\frac{\partial P}{\partial n},\\ 
f_{\rm G}^{n} &=& - \frac{G M}{(R^2+Z^2)^{3/2}}(R\vec{e_R}+Z\vec{e_Z})\cdot
\vec{e}_{\rm n},\\ 
 f_{\rm C}^{n} &=&   V_{\theta} R \vec{e_r}\cdot
\vec{e}_{\rm n},\\
  f_{\rm M}^{n} &=&  f_{\rm M_{\rm p}}^{n} + f_{\rm M_{\theta}}^{n} = (\vec{J}\times\vec{B})\cdot
 \vec{e}_{\rm n}\,.
\end{eqnarray}
where $f_{\rm P}^{s,n}$, $f_{\rm G}^{s,n}$, $f_{\rm C}^{s,n}$, $f_{\rm
M}^{s,n}$, $f_{\rm M_{\rm p}}^{n}$ and  $f_{\rm M_{\theta}}^{n} $ correspond 
respectively to the  pressure gradient, the gravitational, the centrifugal, 
and the magnetic forces, one induced by the poloidal magnetic field component 
and the other by the toroidal one. The indexes $^{s}$, $ ^{n}$ denote 
forces parallel and perpendicular to the poloidal magnetic field line, 
respectively. We have defined the unit vector $\vec{e}_{\rm s}$ that corresponds to
$\vec{B_p}/|B_p|$ in the left panel of Fig.~\ref{Fig8} and to
$\vec{V_p}/|V_p|$ in the right panel of Fig.~\ref{Fig8}, while the
perpendicular unit vector is such that $\vec{e}_{\rm n}\cdot\vec{e}_{\rm
  s}=0$ in Fig.~\ref{Fig9}.  
As the stellar wind is resistive, the flow streamlines do not have to be
parallel to the 
poloidal magnetic field to reach a quasi steady-state, since $\bf{V}_p \times
\vec{B}_p= \eta_m \vec{J}_\theta\neq 0$. 
\subsubsection{MHD disc-driven jets}
\begin{figure*}[t]
{\rotatebox{0}{\resizebox{9cm}{8cm}{\includegraphics{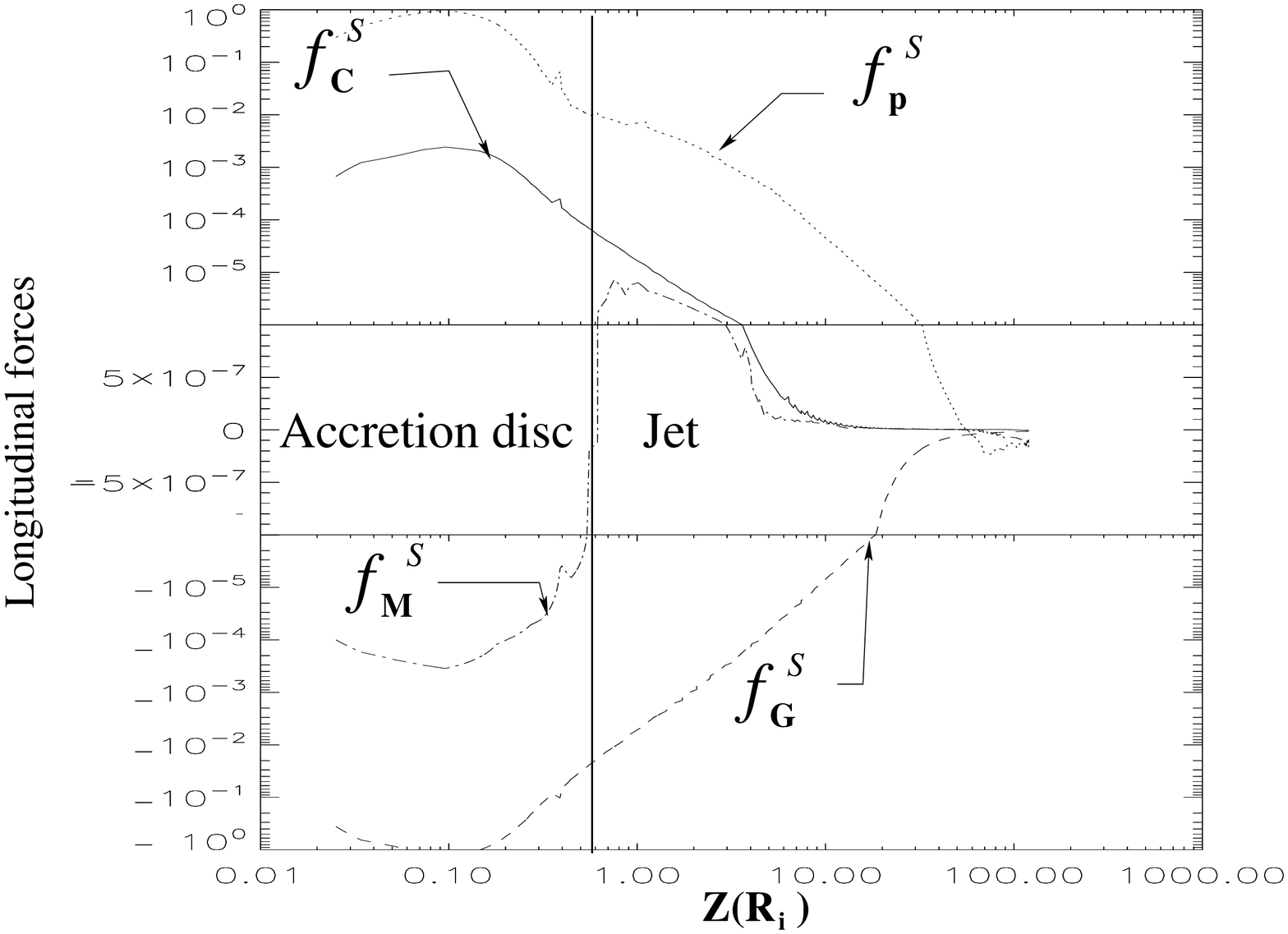}}}}
{\rotatebox{0}{\resizebox{9cm}{8cm}{\includegraphics{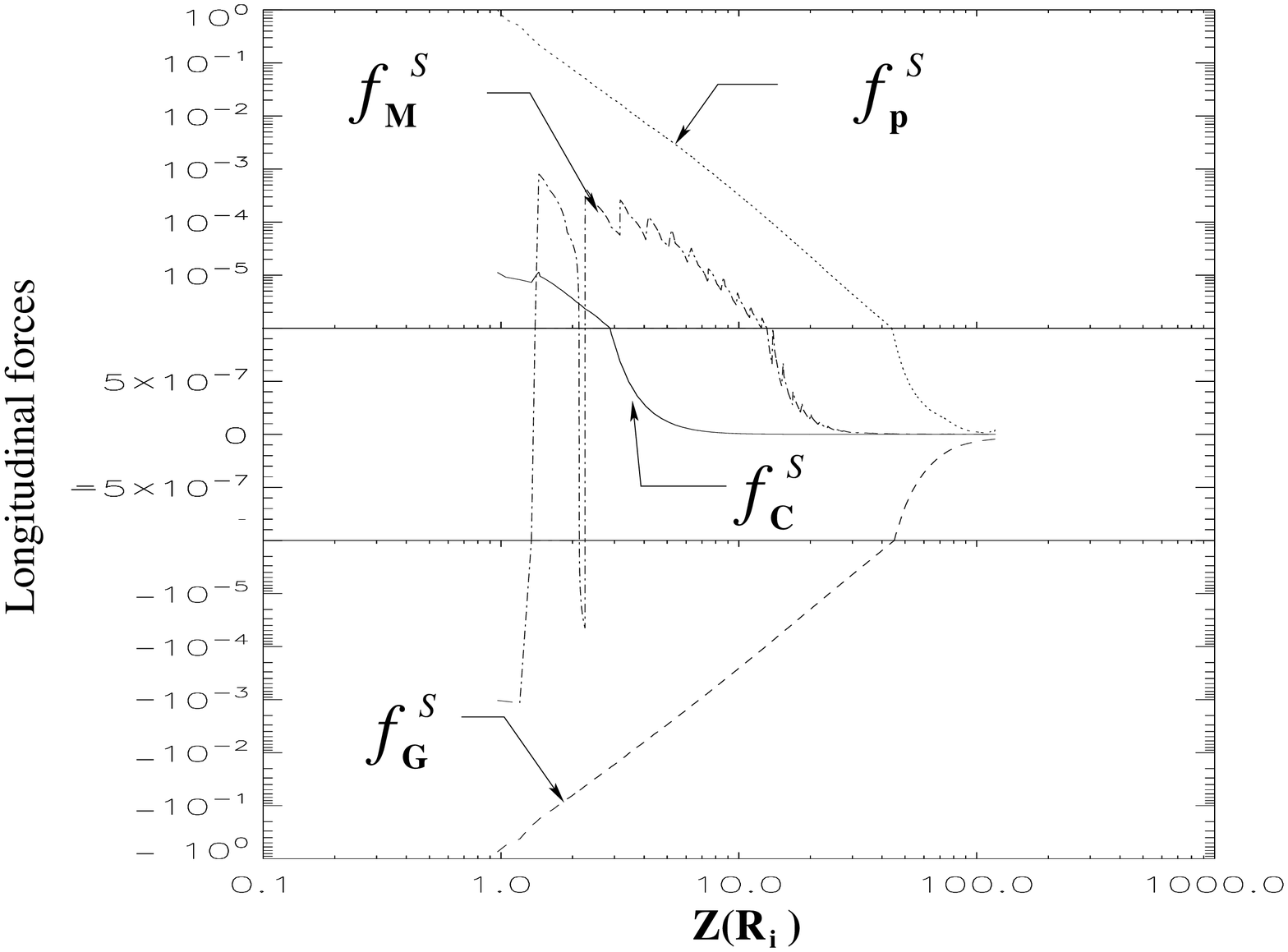}}}}
\caption{Same plots as in Fig.~\ref{Fig8} but for a
  simulation where the
  stellar mass ejection rate is $\dot{M}=10^{-7}M_\odot/yr$.} 
\label{Fig11}
\end{figure*}
The disc-driven jet behaves as in CK04. In particular, 
we find that the mass acceleration encounters two different regimes. In
Fig.~\ref{Fig8}  we see that the vertical  outflow is
lifted from the accretion disc by both the magneto-centrifugal force and the
pressure gradient up to the Alfv\'en surface. Beyond, the poloidal acceleration
is mainly sustained by the pressure  gradient associated with the toroidal
component of the magnetic field. Inside the resistive accretion disc, the
toroidal component of the magnetic field increases because of the differential
rotation of the disc $ \nu_m \partial B_\theta/\partial s \sim  \int_{0}^{s}
{\rm d}s r \vec{B}_p \cdot \vec{\nabla} \Omega$. Conversely, outside
the disc, the  advection of the toroidal
magnetic field balances the differential rotation $ \nabla
\frac{1}{r}(B_\theta \vec{v}_p) = \nabla (\vec{B}_p \Omega)$ exactly
\citep{Ferreira97,CasseFerreira00}.  This change induces a decrease in
$B_{\theta}$ outside the accretion disc. As shown in CK04, this configuration
allows matter below the disc surface to be pinched and to remain in an
accretion regime, while beyond the disc surface, the change of sign of
$ f_{\rm M}^{s}$ enables acceleration of mass along the magnetic field
lines (cf Fig.~\ref{Fig8}). This change in the magnetic poloidal force 
also corresponds to a change in sign of the magnetic torque
($(\vec{J} \times \vec{B})\cdot \vec{B}_p = -(\vec{J} \times
\vec{B})\cdot\vec{B}_{\theta}$ ) leading to the transformation of the MHD Poynting
flux generated by the disc into kinetic energy of the jet material.\\
The cylindrical collimation of the external outflow is induced by the pressure
gradient of the poloidal component of the magnetic field ($f_{M_{\rm P}}^n$ in
Fig.~\ref{Fig9}). In  fact, the magnetic field 
in the disc-driven jet  undergoes an expansion that induces a decrease
in the poloidal magnetic field in the jet compared to the outer  region
(Fig.~\ref{Fig9}). Conversely the pressure gradient of the toroidal magnetic 
field ($f_{M_{\rm \theta}}^n$) acts to decollimate the outflow because the 
value of $B_{\theta}$ is low outside the outflow, and its absolute value 
decreases between the massive part of the disc-driven outflow and the outer 
medium. The magnetic field lines in the massive part of the outflow are 
anchored to the inner part of the accretion disc and extract more angular
momentum than the magnetic lines in the outer medium.
The inner part of the jet is, on the other hand, collimated by the toroidal 
pinching force.

\subsubsection{Stellar wind embedded in a disc-driven jet}\label{Stellarwind}

\begin{figure}
{\rotatebox{0}{\resizebox{8cm}{9.cm}{\includegraphics{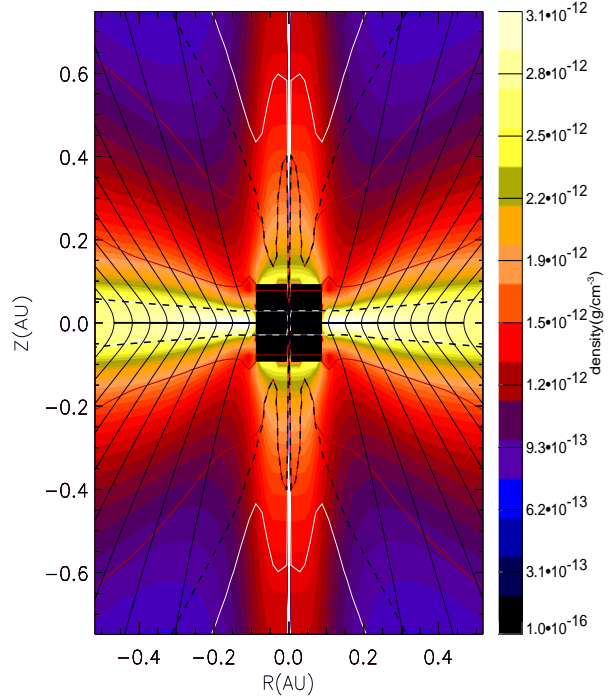}}}}
\caption{ Same density plot as in Fig.~\ref{Fig3} but for a smaller zone around the
  sink. The three critical surfaces are represented by dark lines
  (slow-magnetosonic), dashed lines (Alfv\`en), and white lines
 (fast-magnetsonic). { The size of the sink region
is $R_i=0.1 {\rm AU}$ and the stellar mass is $1 M_{\odot}$.}
 The sink region is represented by a black square 
in the center} 
\label{Fig12}
\end{figure}

\begin{figure*}
\begin{center}
{\rotatebox{0}{\resizebox{8cm}{5.cm}{\includegraphics{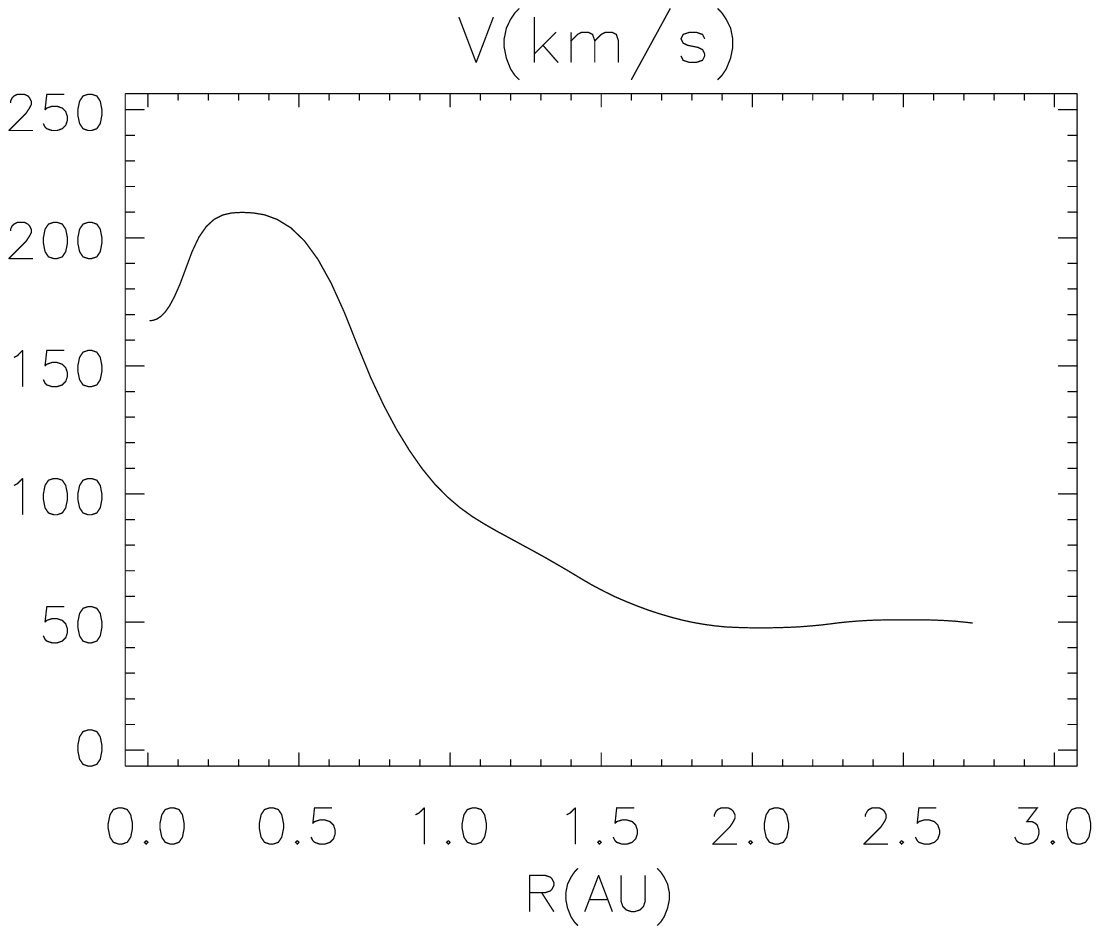}}}}
{\rotatebox{0}{\resizebox{8cm}{5.cm}{\includegraphics{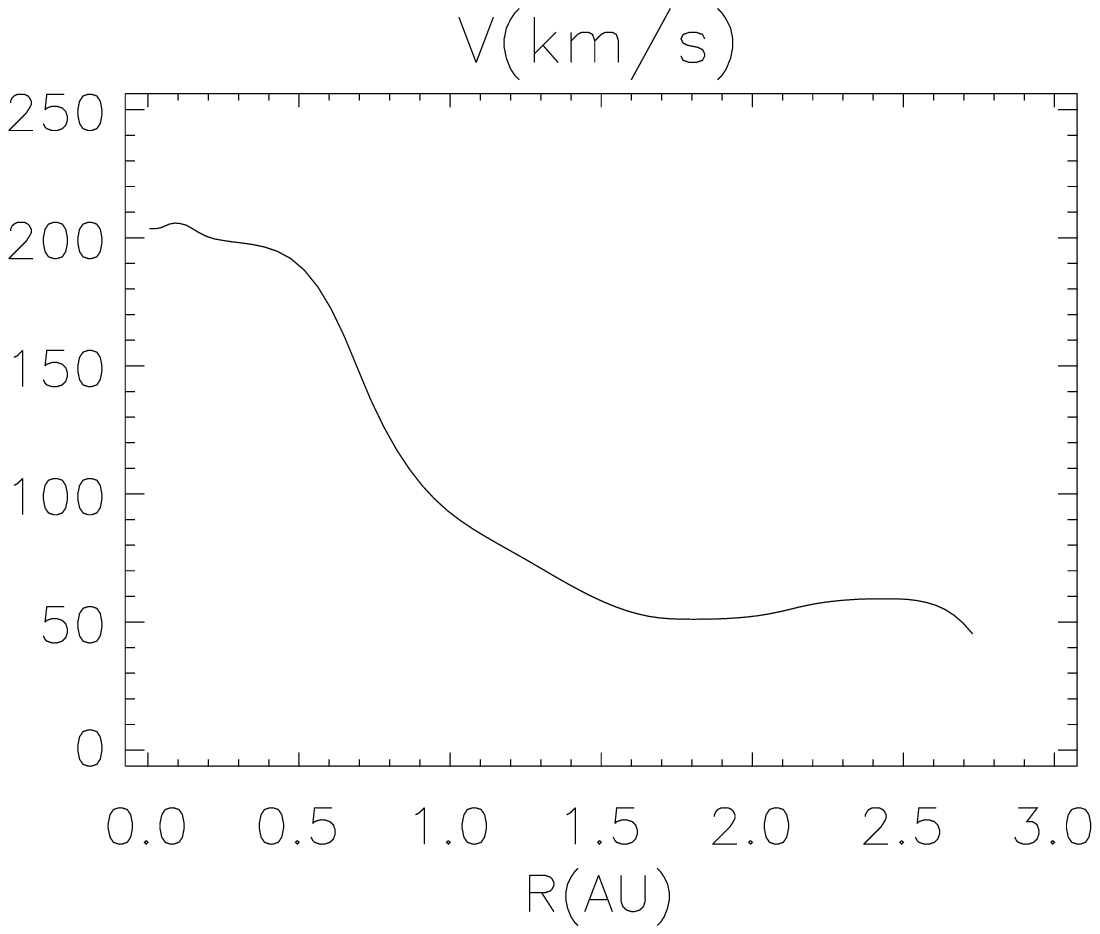}}}}
\end{center}
\caption{The transverse variation of the vertical velocity
 at $Z = 8 AU$, for the simulation with stellar mass loss 
$\dot{M}=10^{-9}M_\odot/yr$, $M_*=M_{\odot}$, and $R_I=0.1 AU$,
{\bf Left:} without any non-ideal effects in the stellar wind but
 with a large amount of thermal energy deposited at the base of the stellar
 outflow ($\delta_{\varepsilon}=10^{-3}$).  {\bf
   Right:} With turbulent viscosity and resistivity in the stellar
 wind and a small amount of thermal energy at the base
 ($\delta_{\varepsilon}=10^{-5}$). The turbulence allows a better thermal
 stellar mass acceleration as velocity becomes higer near the polar axis.} 
\label{Fig13}
\end{figure*}

The stellar wind undergoes a thermal acceleration, as long as the shape of the
jet does not become cylindrical. Incontrast to the disc-driven jet, the
magneto-centrifugal force remains weak along the stellar wind flow. 
The opening angle between magnetic field lines that emerge from the sink region 
remains weak.
In fact the stellar outflow starts to be  collimated by the surrounding
hollow jet induced by the accretion disc.   
This explains the difference from a
model of \citet{Matt&Balick04} where the stellar wind is the  only 
outflow and which is prone to a dipolar expansion.  
In our simulation, we self-consistently describe the acceleration of the 
inner jet in addition to its collimation, something that can be compared
with analytical modeling \citep{STT02, STT04}. Part of both the thermal energy
deposited at  the surface 
of the corona near the polar axis and the energy deposited by the turbulence in 
the stellar wind is transformed into kinetic energy (Fig. \ref{Fig10})
along the  streamline.  We have estimated the amount of Ohmic heating
  released in the stellar outflow in the context of our simulation, namely,
\begin{equation}
P_{Ohmic} = \iiint_V(\eta_mJ^2 - \vec{B}\cdot(\nabla\times\eta_m\vec{J}))dV
\end{equation}
where $V$ stands for the stellar outflow volume. The volume-integrated 
heating represents $35\%$ of the energy released by accretion.
  However, the  streamlines in the stellar wind are 
subject to a larger expansion (relatives to the jet cylindrical radius at
the Alfv\`en surface) than the streamlines anchored to the accretion
disc. 
Therefore, the angular momentum conservation induces a decrease in
$V_\theta$ and $B_\theta$. In the asymptotic region, the magnetic 
field lines become almost  radial so that the projection of the magnetic
pressure gradients along the magnetic field lines  become positive. The
flow undergoes magnetic and thermal acceleration in this region.\\
 The collimation of the inner part of the jet is induced by the thermal pressure
plus the pinching of the toroidal component of the magnetic field. They balance
the centrifugal force and the pressure of the poloidal component of 
the magnetic field (Fig.~\ref{Fig9}). Besides that, the
simulation shows that the inner portion of the stellar jet 
has a deep in density (Fig.\ref{Fig4}) and a peak of velocity around the
axis (Fig.~\ref{Fig13}).  
Thus, all these facts suggest that 
 the very inner part of the outflow, the so-called spine jet, behave as  a 
meridionally-self-similar jet as in \citet{STT02, STT04}. 
We observe that this is a kind of ``hollow" stellar jet 
``thermally" driven and both magnetically and thermally confined inside the 
``hollow" disc jet. This 
result is to be compared to the analysis of CK04 where it is shown that the 
external disc jet is, partially at least, behaving as a  radially self-similar
one.\\
In Fig.~\ref{Fig12}, we display temperature isocontours within
  a small area around the sink region. Thanks to this plot, we can see that
  the magnetic lever arm associated with the various outflow components is
  different. In fact, the disc-driven jet exhibits a magnetic lever arm
  (related to the ratio of the Alfv\`en radius to the magnetic-field line
  foot-point radius) varying approximately from $9$ to $25$, while the magnetic 
  lever arm associated with the stellar wind ranges from $0$ near the axis to 
  several tens, if one considers the foot-point of magnetospheric magnetic-field
  line to be anchored to the star. This last magnetic lever arm value may not 
  be very reliable since we have imposed the size of the sink and thus
  influenced the radial extension of the magnetospheric outflow near the sink.

\subsubsection{Ideal MHD two-components outflows}
\label{IdealJet}
 
An alternative to the presence of turbulence in the stellar wind would
  be to have a mechanism acting near the star corona and transforming a
  part of the accretion energy into thermal energy (see for instance
  Matt \& Pudritz 2005 and reference therein). We performed several simulations 
  without turbulence inside the  stellar wind ($\alpha_w=0$)  and with the 
  value of $\delta_{\varepsilon}$  regularly increased (and thus the amount 
  of thermal energy at the top of the sink). We find that, in our 
  simulations, the maximal
  value of $\delta_{\varepsilon}$ is around $ 10^{-3}$. Beyond that value,
  the pressure above the sink is so high that it disrupts the accretion
  disc structure and prevents the launching of the disc-driven jet. It is
  noteworthy that this value of $\delta_{\varepsilon}$ is linked to very
  high value of thermal energy released in the star corona. 

  Indeed, the reader has to keep in mind that the top boundary of the sink 
  is quite far from the stellar surface (typically $20$ stellar radii) so that 
  if the flow undergoes a spherical expansion with a constant thermal energy 
  flux, the thermal energy deposited in the corona would represent
  $\delta_{\varepsilon}(R_i/R_*)^2$ of the energy released by
  accretion. This amount of thermal energy may then represent a
  significant fraction of the accretion energy. In our calculations done
  without stellar outflow turbulence, we  noticed that the 
  structures fulfilling observational constraints coincide with the highest
  value of $\delta_{\varepsilon}$ allowing disc-driven jet launching
  (typically $10^{-3}$). 

  The resulting two-component outflow is very similar to simulations done with 
  turbulence in the stellar wind (and a very low $\delta_{\varepsilon}$), 
  except the terminal velocity of the stellar outflow as shown in 
  Fig.~\ref{Fig13} where we display the vertical velocity of matter along a 
  radial direction located at $Z=80R_i$. In this figure we can clearly see that 
  the stellar wind prone to turbulent heating is faster than the ideal MHD 
  stellar wind. The poloidal mass acceleration in this zone is very sensitive 
  to thermal heating since magneto-centrifugal is vanishing here. A continuous 
  heating, as generated by Ohmic heating, seems more efficient for accelerating 
  the mass since it is
 ``refilling'' the thermal energy reservoir available for acceleration
   along the flow.  \\
\begin{figure}[t]
\begin{center}
{\rotatebox{0}{\resizebox{7cm}{12cm}{\includegraphics{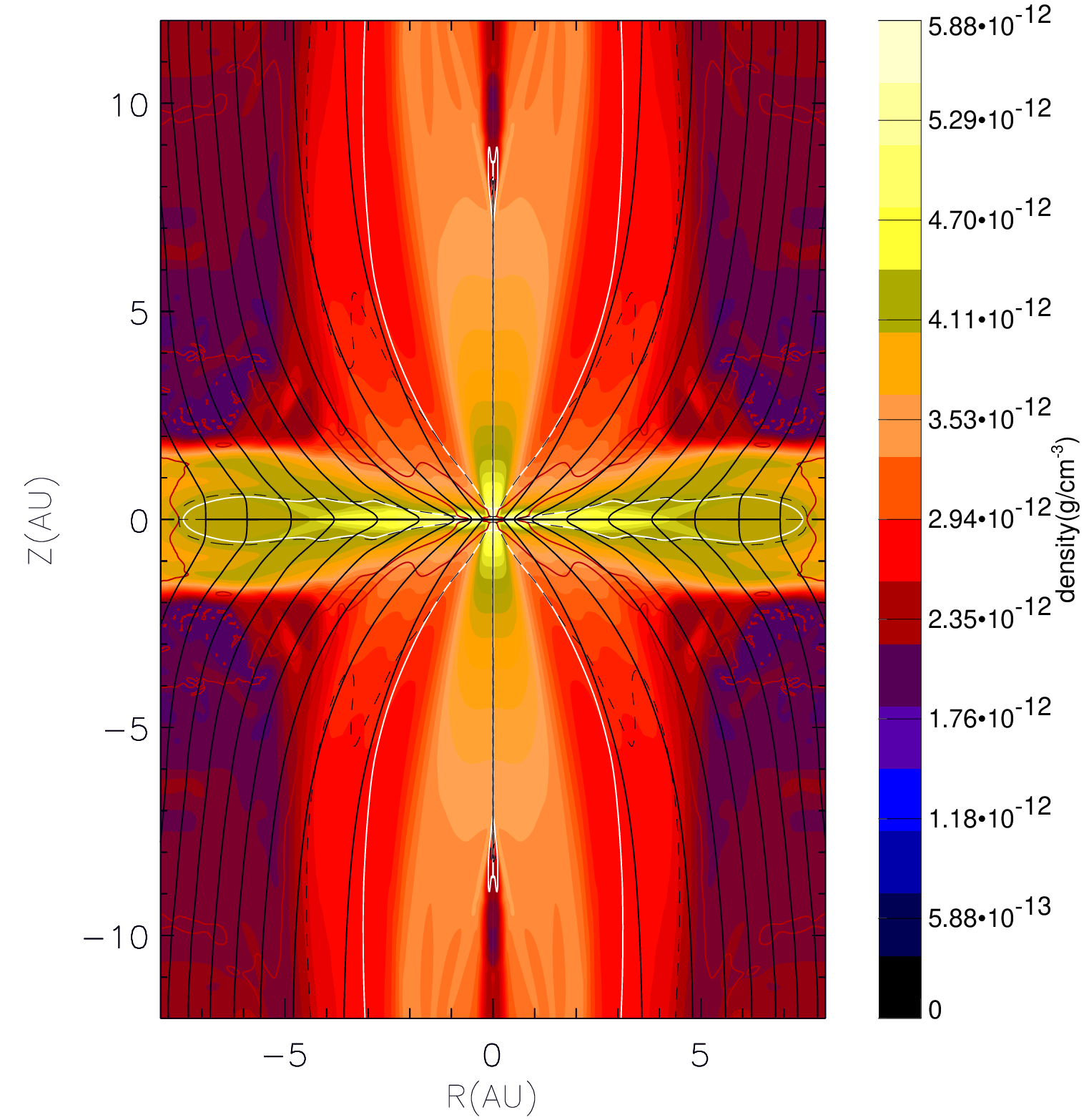}}}}
\end{center}
\caption{Same figure as in Fig.~\ref{Fig1} but with a
  non-ideal stellar wind emitted from the inner region with an ejection mass
  rate $\dot{M}=10^{-7}M_\odot/yr$. The outflow structure is substantially
  modified by the 
  presence of the stellar outflow since its radial extension is two times
  larger than
  in the case with no, or weak,  stellar outflow. { The size of the sink region
is $R_i=0.1 {\rm AU}$ and the stellar mass  $1 M_{\odot}$.}}
\label{Fig14}
\end{figure}

\subsection{Massive stellar winds vs. sun-like mass-loss rate wind 
effects on a disc-driven jet} 
\begin{figure*}
{\rotatebox{0}{\resizebox{16cm}{5cm}{\includegraphics{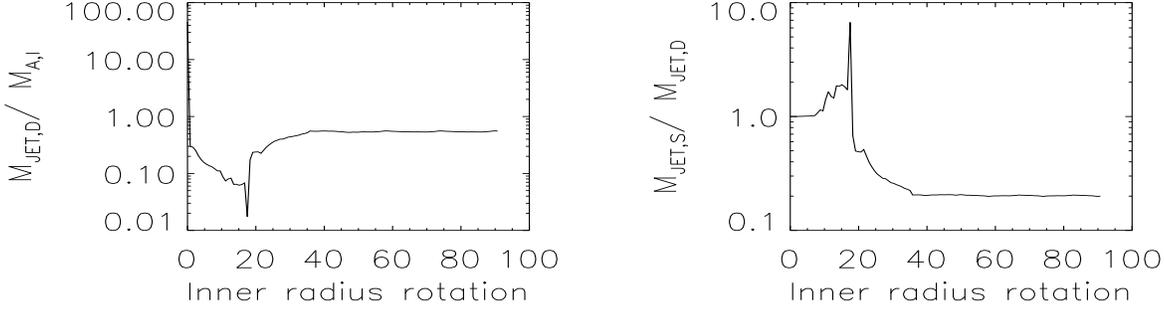}}}}
\caption{Plot of the temporal  evolution of the ejection  mass-loss rate from 
the accretion disc in a) the ratio of 
the stellar mass-loss rate to the ejection  mass-loss rate 
from the accretion disc as a function of time in b)  for the simulation with a 
stellar mass-loss rate of $\dot{M}=10^{-7}M_\odot/yr$.} 
\label{Fig15}
\end{figure*}

\begin{figure*}
{\rotatebox{0}{\resizebox{18cm}{5.5cm}{\includegraphics{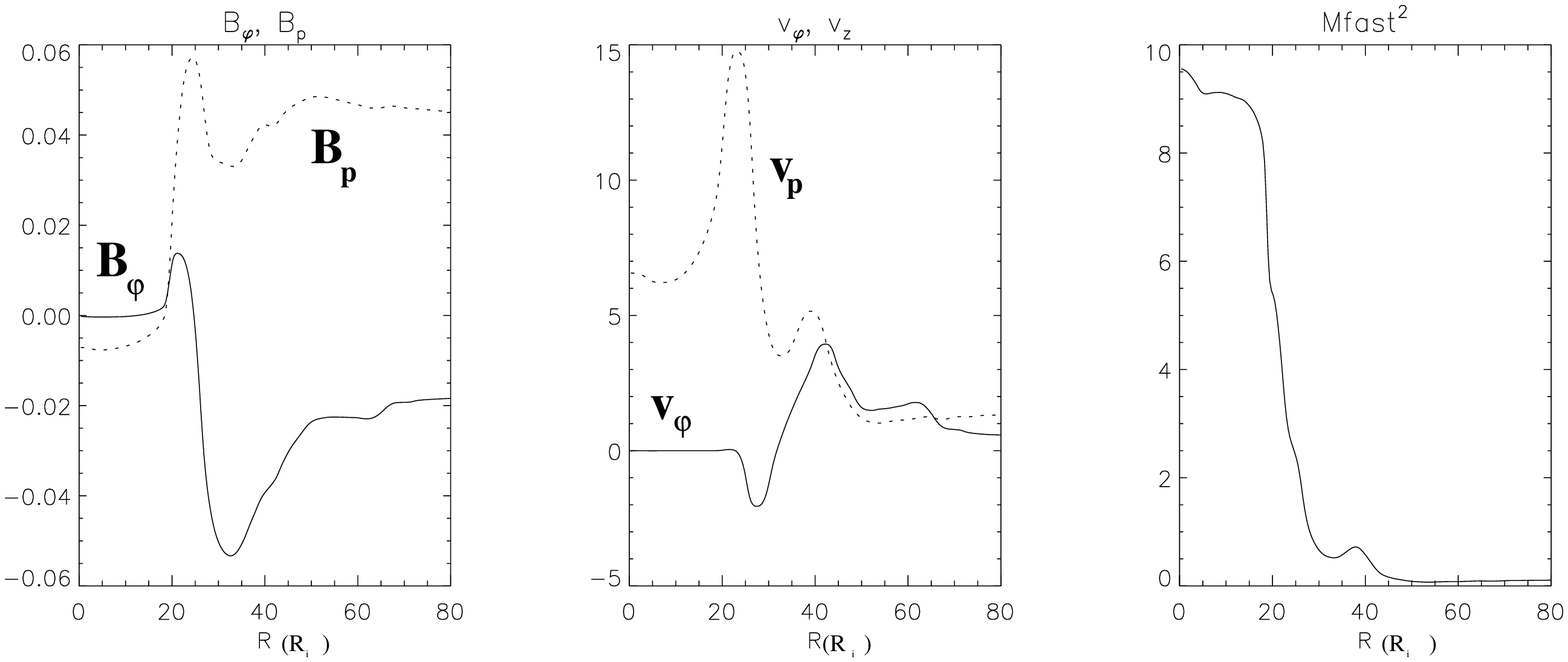}}}}
\caption{The transverse variation of different physical quantities as
  magnetic-field components, velocity components, and the fast-magnetosonic
  Mach number
 at Z = 100, for the simulation with stellar mass loss 
$\dot{M}=10^{-7}M_\odot/yr$}. 
\label{Fig16}
\end{figure*}

In the simulations presented so far, we have seen  that winds
with a mass-loss rate similar to the Sun (up to $10^{-9}M_\odot/yr$)
do not greatly influence the disc outflow since their
general behavior remains similar. However in the case of a massive
stellar jet,  the inner wind may strongly influence the outflow as can be 
seen in a new simulation performed for a stellar-wind mass-loss rate  set to
$10^{-7}M_{\odot}/yr$ (Fig.~\ref{Fig14}).
The radial stellar wind strongly compresses the magnetic field anchored in the 
accretion disc. The enhanced magnetic field bending (even in the external part 
of the accretion disc $R>30$) leads to an increase in the magnetic pinching in 
an extended region of the disc $1<R<30$. Thus the outflow is launched from  
this whole region since the Blandford \& Payne criterion is fulfilled everywhere
\citep{BP82}.  Indeed the magnetic field becomes dynamically dominant in the 
disc corona of this region. The magnetic bending larger than $30^{\circ}$
from the vertical direction leads to a centrifugal force and a thermal
gradient pressure that is more efficient for launching the outflow from the disc 
(Fig.~\ref{Fig11}) as  can be seen in the jet mass loss that reached $0.5$ of 
the accretion rate in the inner part  (Fig.~\ref{Fig15}). However, the amplitude of the 
magnetic force $\vec{e_{\rm p}} \cdot(\vec{J}\times \vec{B_{\theta}})$
projected along the  poloidal
direction becomes weak since the  projection along $\vec{e_{\rm p}}$  of
the magnetic pinching force increases with the expansion of the magnetic
field (see Fig.~\ref{Fig11}).\\
The angular momentum carried away by the stellar outflow now  represents $5\%$ of 
the accreted angular momentum at the inner radius of the accretion disc. 
Regarding the acceleration of the outflow, we can distinguish two regions:
an internal one corresponding to the contribution from the stellar outflow
and an external one coming from the disc-driven
jet. This last component reaches velocities up to $v_z = 15$
(Fig.~\ref{Fig16}). 
The acceleration of this component is thermally and
magneto-centrifugally driven, which is coherent with the larger radial
expansion of the stream 
lines (see Fig.~\ref{Fig11}). In the inner stellar wind,
the flow undergoes  a weak expansion, and its velocity does not exceed $v_z =
6.8$. The acceleration of this component is achieved mainly via the 
thermal pressure, which is expected since the mass density is much higher
than in previous simulations where $\dot{M}=10^{-9}M_{\odot}/yr$. Let us
note that the fast-magnetosonic Mach number remains higher than one for the
whole outflow (Fig.~\ref{Fig16}).

\section{Summary and concluding remarks}

In this paper we present numerical simulations of the
interaction between an  accretion-ejection 
structure launching a disc-driven jet and a stellar wind emitted either from the
central object and/or from its magnetosphere, in particular for the case of YSOs. In our framework, the
accretion disc is near 
equipartition between thermal pressure and magnetic pressure where
turbulence is believed to occur. This turbulence is characterized by a time
and space-dependent anomalous  resistivity and viscosity set by using an
$\alpha$ description. The origin of the turbulence is still unknown but is
not likely to arise from magneto-rotational instability since an
equipartition disc is inconsistent with the development of such instability
\citep{Ogilvie&Livio01}. \\
The properties of both the accretion disc and  outflow were investigated in 
this paper.  In a first stage, we analyzed the contribution of the various
hydrodynamical and magnetohydrodynamical mechanisms   
to the angular momentum transport in the thin accretion disc including, for the 
first time, both anomalous viscosity and resistivity, with a magnetic
Prandtl  number equal to unity.  We demonstrated 
that the MHD Poynting flux associated with the disc-driven jet plays a
major role in the removal of the angular momentum from
the thin accretion disc. The angular momentum radial transport provided by
the anomalous viscosity inside   
the  disc remains weak and contributes only  $1\%$ of the total angular
momentum transport (this value  agrees with analytical estimates
depending on disc thickness and $\alpha$ value). This is consistent with
the viscous torque depending upon the radial derivative of angular velocity, 
while the magnetic torque is mainly controlled by the vertical derivative of the
toroidal component of the magnetic field. In a thin accretion disc, the
toroidal magnetic field varies from zero to an equipartition value on a
disc scale height $\epsilon R, \varepsilon \ll 1$, leading to a much more 
efficient extraction of the rotational energy from the magnetic torque into
the MHD Poynting energy flux feeding the jet.  
Basically, with our simulation and despite the disc viscosity, we have
reproduced very similar results to those obtained by CK04 where the
viscous torque was neglected.\\ 
In the second stage of the present paper, we studied the effects
induced  by the launching of a stellar wind inside the 
 hollow jet arising from the accretion disc.
The stellar outflow is thermally driven by the turbulent viscous and
resistive stresses in a region close to the
polar axis. In order to mimic the non-ideal effects believed to occur
inside the stellar wind, we prescribed anomalous viscosity and
resistivity in the wind region, leading to a turbulent heating of the
plasma near the polar axis.\\ 
We performed various simulations using different stellar mass ejection
rates from the central objects. These stellar ejection-mass rates range
from a Sun-like star ($\dot{M}=10^{-9}M_{\odot}/yr$) to O and B type stars
($\dot{M}=10^{-7}M_{\odot}/yr$).
The influence of the stellar wind on the
dynamics and the structure of jet and the accretion-ejection structure around the
stellar object gets stronger with higher stellar mass ejection rates. As an
example in the simulations where stellar mass ejection rates are
$\dot{M}=10^{-9}M_{\odot}/yr$, we obtained a very similar disc-driven jet to the 
one in CK04. The only difference lies in the presence of the internal, fast, hot 
plasma coming from the central object. In this simulation, the ejection-mass 
rate in the disc-driven jet is similar to the one in CK04, on the order of
$15\%$ of the inner disc accretion mass rate, while the stellar ejection
mass rate represents $1\%$  of the total mass loss in the outflow. 
 
In our simulations, the collimation of the stellar outflow takes place once
the  
jet from the accretion disc is launched and has reached a significant spatial 
extension. Its collimation is provided mainly by pinching the toroidal magnetic 
field in equilibrium with the thermal pressure gradient. Conversely the 
collimation of the jet from the accretion disc is induced by the poloidal 
magnetic field pressure gradient balancing the centrifugal force. 
Furthermore, in all the simulation the stellar wind keeps having a more or less
conical expansion up to
the asymptotic region where the disc-driven jet acts to collimate the
stellar flow into a cylindrical flow.
 
These important results are  self-consistently obtained, in contrast to
simulations of 
\citet{Bogovalov&Tsinganos05} where a relativistic wind was collimated by a
 jet but which did not consider either the accretion disc, the jet
launching, or the stellar wind acceleration. Indeed in our model,
we  self-consistent describe the launching and the collimation of disc and 
stellar wind. In particular, our simulations completely describe both the stellar flow 
acceleration
(the stellar flow is injected with sub-Alfv\`enic velocity) and the
 launching mechanism of the jet from the accretion disc. 

We have shown that the contribution of  non-ideal MHD 
mechanisms in the 
acceleration of the stellar outflow can be significant since turning on
this dissipative mechanism leads, for instance, to higher terminal
velocities of the stellar jet-collimated flow. Our prescription of these
dissipative mechanisms is of course subject to improvements, but our goal
was to show that they  enable an increase in the efficiency
of both the thermal and  magnetic acceleration of the stellar wind.
The turbulence may be produced by the interaction of the stellar wind with
the disc-driven jet. Moreover, as in the solar wind, the turbulence in 
the stellar wind may also have a stellar origin and/or  a possible connection
 to the accretion occurring near the surface of the star. In this scenario,
 a part of the energy released by accretion is carried away in the wind by
 outwardly propagating Alfv\'en waves inducing  turbulence. This scenario
 is a variant of models where a significant part of the accretion
 energy is converted into thermal energy in the star corona (see
 Matt \& Pudritz 2005 and references therein). By performing
 simulations with no stellar wind turbulence but with a large amount of
 thermal energy at the base of the wind, we found quite similar result
except for the velocity field, the resistive continuous heating of the
stellar wind being more efficient in providing higher velocity. It is noteworthy 
that, similarl to models depositing thermal energy near the stellar corona, 
the amount of energy released by the turbulent heating is a significant 
fraction of the accretion energy (in the particular case of our simulations, 
it represents near $35\%$ of the accretion energy). 
Note also that similar double-layer jets were found by Koide and
collaborators in  various
simulations (e.g. Koide et al. 1998; Koide et al. 1999) but those 
simulations were devoted to
rapidly variable jets (with only a few disc rotations) and not to steady
structures.\\
We performed simulations with higher stellar ejection-mass  rate,
typically with $\dot{M}=10^{-7}M_{\odot}/yr$ (compatible with O-B type
stars). The increase in the stellar mass-loss rate induces
a faster and larger expansion of the jet. Indeed the enhanced pressure
provoked by the stellar wind tends to bend the disc magnetic field lines over a
larger radial extension, leading to a larger disc-driven jet. The
corresponding disc-driven jet mass-ejection rate is much higher than in
previous simulations since it reaches $50\%$ of the disc inner accretion
rate (stellar ejection-mass rate on the order of $10\%$).  
The simulations give a quantitative 
threshold beyond which the stellar jet gives a significant extra expansion
of the disc jet. Typically 
a mass loss rate from the star on the order of $\dot{M}=10^{-7}M_{\odot}/yr$
gives a factor two in the radial
expansion of the disc-driven jet. Although the total jet remains small in
cross section, as in CK04, and compared to self-similar disc wind models. Moreover
the stellar jet play a very important role, also, in the formation of the disc
wind. 
We verify that,  part of the external disc wind may look quasi
radially-self-similar in nature, the  
most inner part of the stellar wind is quasi meridionally self-similar.
In our work we neglected all radiative losses coming from the
central star or from the plasma itself. The implementation of these terms
and the study of their impact on the outflow structure is postponed to
future work.
\begin{acknowledgements}
ZM thanks Henk Spruit, Andrea Merloni and Dimitrios Giannios for many valuable 
suggestions.
ZM is grateful for the hospitality of the Garching group. Part of this 
research was supported by European FP5 RTN "Gamma Ray Burst: An Enigma and a 
Tool".
ZM thanks Kanaris Tsinganos for many valuable 
suggestions. ZM thanks Claudio Zanni for many valuable 
suggestions. Finally, FC would like to thank Sylvie Cabrit for many helpful
remarks and advises.
\end{acknowledgements}

\bibliographystyle{aa}

\end{document}